\documentclass[aps,prfluids,longbibliography,reprint,superscriptaddress]{revtex4-2}

\usepackage{graphicx}
\usepackage{amsmath,amssymb}
\usepackage{hyperref}
\hypersetup{colorlinks=true,allcolors=blue}
\newcommand{\bfu}{\mathbf{u}}
\newcommand{\bfw}{\boldsymbol{\omega}}
\newcommand{\curl}{\nabla\times}
\newcommand{\grad}{\nabla}
\newcommand{\Div}{\nabla\cdot}
\newcommand{\E}{E}
\newcommand{\Z}{Z}
\newcommand{\D}{D}
\newcommand{\PiC}{\Pi}
\newcommand{\Cfour}{\mathcal{C}_{4}}
\newcommand{\Cp}{\mathcal{C}_{p}}
\newcommand{\Zspec}{\mathcal{Z}}

\newcommand\HL[1]{{\color{black}#1}}

\begin{document}

\title{\HL{Curvature-weighted spectra anticipate} dissipation \HL{peaks} in decaying three-dimensional turbulence}

\author{Satori Tsuzuki}\email{tsuzukisatori@g.ecc.u-tokyo.ac.jp}
\affiliation{Research Center for Advanced Science and Technology, The University of Tokyo}

\date{\today}

\begin{abstract}
We investigate the robustness of a curvature-weighted spectral precursor to dissipation in freely decaying three-dimensional incompressible turbulence. Building on our recent work in \emph{Physical Review Fluids} on the Taylor--Green vortex, we analyze direct numerical simulations using the \HL{shell-summed} curl-of-vorticity spectrum\HL{, denoted here by $\Cfour(k,t)$ and} equivalent to a $k^4$-weighted energy spectrum \HL{in the modal incompressible sense}. Extending the study across multiple initial conditions---multi-mode ABC flows, a randomized low-wavenumber ABC field, the Taylor--Green vortex, and the Kida--Pelz flow---we find a consistent temporal ordering: the characteristic time associated with the advance and saturation of the peak wavenumber of \HL{$\Cfour(k,t)$} precedes the dissipation-peak time, which in turn precedes the characteristic time associated with the peak scale of the nonlinear energy-flux spectrum. We further probe \HL{Reynolds-number and scale-separation effects using} Taylor--Green \HL{simulations at additional viscosities}: the precursor \HL{ordering persists when adequate scale separation and resolution are maintained, but can change in the low-$R_\lambda$/limited-scale-separation regime}. Throughout, we use explicit inspection of curvature-weighted spectra to distinguish physical peak evolution from cutoff-proximate artifacts. These results \HL{support robustness over the deterministic decaying-flow initial conditions examined here} and clarify the practical role of \HL{Reynolds number, scale separation, and resolution when using} curvature-weighted spectral \HL{diagnostics} in decaying turbulence.
\end{abstract}

\maketitle

\section{Introduction}
Predicting \emph{when} a three-dimensional (3D) turbulent flow reaches its dissipation peak is central to both theory and practice.
In freely decaying turbulence, the dissipation rate $\varepsilon(t)$ exhibits a pronounced transient maximum that reflects the formation of the smallest dynamically active scales.
Accurate anticipation of this event is relevant not only for physical interpretation of nonequilibrium cascade development, but also for adaptive strategies in simulations and experiments, such as output scheduling and dynamic resolution control.
Despite decades of progress, reliable \emph{precursors} that provide actionable early warning of the dissipation surge remain limited, especially in transient, non-stationary settings.

\HL{In freely decaying turbulence, the build-up of small-scale activity, the timing of the dissipation maximum, and the redistribution of spectral content are not generally in instantaneous equilibrium. This makes dissipation-peak timing a useful diagnostic problem in its own right: physically, because it probes how rapidly nonlinear transfer populates dissipative scales, and practically, because it can help identify when a decaying flow enters its most intense small-scale stage. The present work therefore studies whether a simple spectral marker based on standard shell-summed quantities can consistently anticipate the dissipation peak across several canonical decaying initial conditions with different symmetry, helicity content, and transient organization.}

A standard spectral description of turbulent transients relies on the energy spectrum $\E(k)$, the enstrophy spectrum $\Z(k)\propto k^2\E(k)$, and the cumulative nonlinear flux \HL{$\PiC(q)$ \cite{Frisch1995,AlexakisBiferale2018}.
Recent work on non-equilibrium dissipation and unsteady turbulence has emphasized that the normalized dissipation coefficient and the relation between transfer and dissipation can evolve substantially during transients, even in otherwise canonical flows \cite{Vassilicos2015,Valente2014_PRE_Imbalance,GotoVassilicos2015,BosRubinstein2017,Zheng2023JFM,Kitamura2025JFM}.
We therefore interpret the present precursor problem not only as a spectral-timing question, but also as part of the broader problem of how small-scale dissipation becomes organized during non-stationary decay.}
In decaying flows, however, these observables do not always serve as sharp predictors of the imminent dissipation peak.
For instance, inertial-range development in $\E(k)$ can be gradual and sensitive to finite-resolution and finite-time effects \cite{Brachet1984,DeBonis2013_TGV_NASA_TM}.
Flux-based indicators are more directly tied to interscale transfer, but the flux peak scale can exhibit an appreciable lag relative to small-scale formation and dissipation \cite{Cardesa2015_PoF_FluxTime,Valente2014_PRE_Imbalance}.
In parallel, real-space visualizations (e.g., $Q$-criterion isosurfaces) often reveal morphological transitions from large-scale coherent structures to filament-dominated states \cite{HuntWrayMoin1988,ChongPerryCantwell1990}, but translating such observations into quantitative early-warning metrics is nontrivial.

\HL{Although the present study is formulated in spectral space, it is closely related in spirit to scale-by-scale descriptions based on the K\'arm\'an--Howarth equation. In that framework, inertial transfer and viscous effects can be distinguished across scales, and the approach to the dissipation maximum may be viewed as a transient reorganization of the balance between these contributions \cite{Lundgren2002,Kitamura2025JFM}. Our curvature-weighted diagnostic does not replace this budget analysis; rather, it provides a compact spectral indicator of when vorticity-gradient content has advanced sufficiently far toward small scales that the ensuing dissipation peak is near.}

In our recent Letter on the decaying Taylor--Green vortex (TGV)~\cite{TsuzukiTGVPrecursorPRF}, we introduced a small-scale weighted diagnostic based on the spectrum of the curl of vorticity.
With $\bfw=\curl\bfu$, incompressibility $\nabla\!\cdot\!\bfu=0$ implies
$\curl\bfw = -\nabla^2\bfu$, \HL{so that the shell-summed curl-of-vorticity spectrum},
\HL{
\begin{equation}
\Cfour(k,t) \equiv \frac12 \sum_{\boldsymbol{\kappa}\in\mathcal{S}_k}
\left|\widehat{\curl\bfw}(\boldsymbol{\kappa},t)\right|^2,
\label{eq:intro_curlw2_equiv}
\end{equation}
}
\HL{is a curvature-weighted spectrum that corresponds to the modal weight $|\boldsymbol{\kappa}|^4|\hat{\bfu}|^2$ and is commonly represented by the shell-center shorthand $\Cfour(k,t)\simeq k^4\E(k,t)$}.
The corresponding peak wavenumber,
\HL{
\begin{equation}
k_{\mathrm{peak}}\!\left[\,\Cfour\,\right](t)
\;\equiv\;
\arg\max_{k\ge 1}\Big\{\,\Cfour(k,t)\,\Big\},
\label{eq:intro_kpeak_def}
\end{equation}
}
was found to advance rapidly to intermediate-small scales and then level off \emph{before} the dissipation rate
$\varepsilon(t)=\sum_k 2\nu k^2 \E(k,t)$ reaches its maximum.
Moreover, the characteristic time associated with this rapid advance precedes the time associated with the largest peak scale of \HL{$|\PiC(q)|$}, yielding the robust ordering $t_k<t_\varepsilon<t_\Pi$ across $256^3$--$1024^3$ resolutions for the TGV \cite{TsuzukiTGVPrecursorPRF}.
This behavior is physically consistent with the $k^4$ weighting: it accentuates the incipient formation of high-curvature sheets and tightly wound tubes, where viscosity ultimately acts most strongly.

A key open question, however, is whether this precursor \HL{persists} across substantially different initial conditions.
The TGV is highly symmetric and has served as a canonical benchmark for the creation of small scales \cite{TaylorGreen1937,Brachet1984,Pereira2021PRF}, but its symmetry may constrain the route to turbulence.
To establish broader relevance, it is essential to test initial conditions that differ in topology, helicity content, and symmetry breaking.
Equally important, the $k^4$ weighting in Eq.~\eqref{eq:intro_curlw2_equiv} can amplify high-wavenumber tails; at finite resolution, this can lead to spurious peak picking near the spectral cutoff unless detection robustness is carefully assessed.
Such issues are typically mild for lower-order weights (e.g., $k^2\E(k)$), but become non-negligible for curvature-biased measures such as \HL{$\Cfour(k,t)$}.

In this work, we extend the precursor framework to a diverse suite of freely decaying 3D flows under a unified numerical and post-processing protocol.
We perform pseudo-spectral direct numerical simulations at viscosity $\nu=10^{-3}$ and base resolution $512^3$ for five distinct initial conditions:
(i) a multi-mode Arnold--Beltrami--Childress (ABC) flow, (ii) an asymmetric multi-mode ABC flow, (iii) an ABC flow augmented by a random-phase low-$k$ perturbation, (iv) the TGV (as a reference), and (v) the Kida--Pelz flow, a classical configuration known to generate intense small-scale activity \cite{Kida1985,Boratav1994,Brachet1984}.

To assess peak-detection reliability in a stringent setting, we additionally perform a $1024^3$ Kida--Pelz simulation, which provides a high-resolution reference and is used as the basis for extracting characteristic times for that initial condition. Across all initial conditions, we observe a consistent temporal ordering in which the characteristic time associated with the rapid advance of \HL{$k_{\mathrm{peak}}\!\left[\,\Cfour\,\right]$} precedes the dissipation-peak time, which in turn precedes the time associated with the largest peak scale of the nonlinear flux:
\begin{equation}
t_k \;<\; t_\varepsilon \;<\; t_\Pi.
\label{eq:intro_ordering}
\end{equation}
We further show that the $k^4$ weighting can, at moderate resolution, produce cutoff-proximate peak locking in peak-based metrics (most clearly for the $512^3$ Kida--Pelz run). Throughout this paper we therefore accompany peak-based results with explicit inspections of \HL{$\Cfour(k,t)$} to verify that the detected maximum lies away from the analysis cutoff. When a cutoff-proximate peak is observed at $512^3$, we treat it as a resolution warning and rely on higher-resolution reference data to extract characteristic times (notably the $1024^3$ Kida--Pelz run).

\HL{A further diagnostic question is whether the observed precursor behavior is specific to the curl-of-vorticity spectrum, or whether it is merely a generic consequence of applying increasingly strong high-wavenumber weights to the energy spectrum. We therefore supplement the cross-initial-condition study with a comparison of the peak-wavenumber trajectories of high-wavenumber-weighted spectra with exponents $p=4,6$, and $8$ (Appendix Fig.~\ref{fig:appendix-highp}). The purpose of this comparison is not to claim mathematical uniqueness of $p=4$, but to assess whether the curl-of-vorticity/palinstrophy-related exponent provides a physically interpretable and practically stable member of this broader family.}

The remainder of the paper is organized as follows.
Section~\ref{secII:GovEqNumMET} summarizes the governing equations, numerical method, and spectral post-processing conventions, including the definitions of $\E(k)$, $\D(k)=2\nu k^2\E(k)$, the transfer $T(k)$, and the cumulative flux \HL{$\PiC(q)$}.
Section~\ref{secIII:InitCondSimSet} describes the set of initial conditions and simulation parameters.
Section~\ref{secIV:diagnostics} introduces the characteristic times $(t_k,t_\varepsilon,t_\Pi)$ and details the robustness checks for peak detection specific to the $k^4$-weighted diagnostic.
Section~\ref{secV:results} presents the main comparative results across initial conditions, including a summary table of $(t_k,t_\varepsilon,t_\Pi)$ and the corresponding lead/lag times.
Section~\ref{secVI:discussion} discusses physical interpretation and limitations, and Section~\ref{secVII:conclusions} concludes with implications and outlook.

\section{Governing equations and numerical method} \label{secII:GovEqNumMET}
\subsection{Governing equations}
We consider freely decaying, incompressible turbulence governed by the 3D Navier--Stokes equations in a periodic box of size $L=2\pi$:
\begin{align}
\partial_t \bfu + (\bfu\!\cdot\!\grad)\bfu &= -\grad p + \nu \grad^2\bfu, 
\label{eq:NS}
\\
\Div\bfu &= 0,
\label{eq:incomp}
\end{align}
where $\bfu(\mathbf{x},t)$ is the velocity, $p(\mathbf{x},t)$ is the kinematic pressure, and $\nu$ is the kinematic viscosity.
The vorticity is $\bfw=\curl\bfu$.
We use the volume average $\langle \cdot \rangle \equiv (2\pi)^{-3}\int_{[0,2\pi)^3}(\cdot)\,d\mathbf{x}$.
The kinetic energy and dissipation rate are
\begin{equation}
K(t) \equiv \frac12\langle |\bfu|^2\rangle,
\qquad
\varepsilon(t) \equiv -\frac{dK}{dt} = \nu \langle |\grad\bfu|^2\rangle.
\label{eq:K_eps_def}
\end{equation}

\subsection{Pseudo-spectral direct numerical simulations}
We solve Eqs.~\eqref{eq:NS}--\eqref{eq:incomp} using a massively parallel pseudo-spectral method with FFTW--MPI slabs and a real-to-complex layout \cite{FFTW3}.
The velocity field is represented by a Fourier series
\begin{equation}
\bfu(\mathbf{x},t) \;=\; \sum_{\boldsymbol{\kappa}} \hat{\bfu}(\boldsymbol{\kappa},t)\,e^{\,i\boldsymbol{\kappa}\cdot\mathbf{x}},
\qquad
\boldsymbol{\kappa}=(\kappa_x,\kappa_y,\kappa_z)\in\mathbb{Z}^3,
\label{eq:Fourier_series}
\end{equation}
on an $N^3$ collocation grid.
Spectral derivatives are computed exactly, e.g.,
$\partial_j u_i \leftrightarrow i\kappa_j \hat{u}_i$.
The nonlinear term
\begin{equation}
\mathbf{N}(\mathbf{x},t) \;\equiv\; (\bfu\!\cdot\!\grad)\bfu
\label{eq:N_def}
\end{equation}
is assembled in physical space from $u_i$ and the spectral derivatives, transformed to Fourier space, and then projected to enforce incompressibility.
In Fourier space, the evolution for each mode can be written as
\begin{equation}
\partial_t \hat{\bfu}(\boldsymbol{\kappa},t)
\;=\;
-\widehat{\mathbf{N}}_{\perp}(\boldsymbol{\kappa},t)
-\nu |\boldsymbol{\kappa}|^2 \hat{\bfu}(\boldsymbol{\kappa},t),
\label{eq:spectral_evolution}
\end{equation}
where $\widehat{\mathbf{N}}_{\perp}(\boldsymbol{\kappa},t)
=\mathbf{P}(\boldsymbol{\kappa})\,\widehat{\mathbf{N}}(\boldsymbol{\kappa},t)$,
and $\mathbf{P}$ is the Leray projector
\begin{equation}
\mathbf{P}(\boldsymbol{\kappa})
=
\mathbf{I}-\boldsymbol{\kappa}\boldsymbol{\kappa}^\top/|\boldsymbol{\kappa}|^2,
\qquad
(\boldsymbol{\kappa}\neq\mathbf{0}),
\label{eq:projector}
\end{equation}
which removes pressure and enforces $\boldsymbol{\kappa}\!\cdot\!\hat{\bfu}=0$ \cite{DomaradzkiRogallo1990,AlexakisBiferale2018}.

Dealiasing is performed by the standard two-thirds rule in a \emph{box} sense during time marching: Fourier coefficients are set to zero whenever any component satisfies $|\kappa_i|>N/3$.
Time integration uses a classical fourth-order Runge--Kutta scheme with an integrating factor for the viscous term \cite{PattersonOrszag1971,CanutoBook}.
Velocity and vorticity snapshots are written at regular intervals for post-processing; per-rank storage enables parallel I/O and exact reconstruction consistent with the FFTW--MPI decomposition.

\subsection{Isotropic spectra, transfer, and flux} \label{subsec:IsoSpecTransFlux}
From saved snapshots we compute isotropic shell \HL{sums} of the energy spectrum $\E(k,t)$, the enstrophy spectrum \HL{$\Zspec(k,t)$}, the dissipation spectrum $\D(k,t)$, and the \HL{shell-summed} curl-of-vorticity spectrum \HL{$\Cfour(k,t)$}.
We denote the wavevector magnitude by $\kappa \equiv |\boldsymbol{\kappa}|$ and define \HL{integer shells by}
\begin{equation}
k \equiv \mathrm{round}(\kappa),
\qquad
\mathcal{S}_k \equiv \{\boldsymbol{\kappa}:\mathrm{round}(|\boldsymbol{\kappa}|)=k\}.
\label{eq:shell_def}
\end{equation}
\HL{With the real-to-complex plane weights denoted by $w_{\boldsymbol{\kappa}}$, the} shell-summed energy spectrum is defined \HL{as}
\HL{
\begin{equation}
\E(k,t)
=
\frac12 \sum_{\boldsymbol{\kappa}\in\mathcal{S}_k}
w_{\boldsymbol{\kappa}}
\bigl|\hat{\bfu}(\boldsymbol{\kappa},t)\bigr|^2,
\qquad
\sum_k \E(k,t)=K(t).
\label{eq:Ek_def}
\end{equation}
}
\HL{The shell-summed enstrophy spectrum is}
\HL{
\begin{equation}
\Zspec(k,t)
=
\frac12 \sum_{\boldsymbol{\kappa}\in\mathcal{S}_k}
w_{\boldsymbol{\kappa}}
\bigl|\hat{\bfw}(\boldsymbol{\kappa},t)\bigr|^2
=
\frac12 \sum_{\boldsymbol{\kappa}\in\mathcal{S}_k}
w_{\boldsymbol{\kappa}}|\boldsymbol{\kappa}|^2
\bigl|\hat{\bfu}(\boldsymbol{\kappa},t)\bigr|^2.
\label{eq:Zk_def}
\end{equation}
}
\HL{The dissipation spectrum is evaluated consistently from the calibrated velocity spectrum as}
\HL{
\begin{equation}
\D(k,t)=2\nu k^2\E(k,t),
\qquad
\varepsilon(t)=\sum_k \D(k,t),
\label{eq:Dk_def}
\end{equation}
}
\HL{where $k$ denotes the shell-center wavenumber used in the binned spectra.
Similarly, the shell-summed curl-of-vorticity spectrum is defined by}
\HL{
\begin{equation}
\Cfour(k,t)=
\frac12 \sum_{\boldsymbol{\kappa}\in\mathcal{S}_k}
w_{\boldsymbol{\kappa}}
\left|\widehat{\curl\bfw}(\boldsymbol{\kappa},t)\right|^2
=
\frac12 \sum_{\boldsymbol{\kappa}\in\mathcal{S}_k}
w_{\boldsymbol{\kappa}}|\boldsymbol{\kappa}|^4
\bigl|\hat{\bfu}(\boldsymbol{\kappa},t)\bigr|^2,
\label{eq:curlw2_equiv_methods}
\end{equation}
}
\HL{where the last equality uses incompressibility.
Because shells are formed by integer binning, the frequently used relations}
\HL{
\begin{equation}
\Zspec(k,t)\simeq k^2\E(k,t),
\qquad
\Cfour(k,t)\simeq k^4\E(k,t)
\end{equation}
}
\HL{should be understood as shell-center shorthand expressions rather than exact identities for the binned spectra.
This distinction is important for high-wavenumber-weighted diagnostics, where shell-center and modal-weighted definitions can differ slightly inside a finite-width shell}.

To quantify interscale transfer we compute the transfer function directly from modal transfers in Fourier space:
\begin{equation}
T(k,t)
  = - \sum_{\boldsymbol{\kappa}\in\mathcal{S}_k}
      {\mathfrak R}\!\left\{
        \hat{\bfu}^{\ast}(\boldsymbol{\kappa},t)\cdot
        \widehat{\mathbf{N}}_{\perp}(\boldsymbol{\kappa},t)
      \right\},
\label{eq:Tk_def}
\end{equation}
where $\mathbf{N}=(\bfu\!\cdot\!\grad)\bfu$ and $\widehat{\mathbf{N}}_{\perp}$ is defined in \eqref{eq:spectral_evolution}--\eqref{eq:projector}.
\HL{To avoid notational conflict with the kinetic energy $K(t)$, the} cumulative flux is \HL{written with cutoff variable $q$:}
\HL{
\begin{equation}
\PiC(q,t) \;=\; -\sum_{m\le q} T(m,t),
\label{eq:Pi_def}
\end{equation}
}
so that \HL{$\PiC(q_{\max},t)\approx0$} and $\sum_k T(k,t)\approx 0$ provide discrete conservation diagnostics (up to round-off).

For the post-processing of isotropic spectra and transfer/flux, we apply a \emph{spherical} two-thirds analysis mask and record it as a binary column \texttt{mask} in the spectra output; unless stated otherwise, all shell sums, peak searches, and plots are restricted to shells with \texttt{mask}$=1$.
The mask keeps only modes with $\kappa\le K_{\mathrm{cut}}\equiv \lceil (2/3)K_{\max}\rceil$ (with $K_{\max}=N/2$) and therefore differs from the \emph{box} two-thirds cutoff enforced during time marching.
Because FFTW stores only half of the spectrum in the real-to-complex layout, contributions are weighted so that the $k_z=0$ and Nyquist planes carry a unit weight while the remaining planes carry weight~$2$, ensuring exact counting of conjugate pairs.
With our FFT normalization, $\sum_k \E(k,t)$ is equal to the volume-averaged kinetic energy $K(t)$.
We accumulate shell sums using Kahan summation to reduce floating-point cancellation \cite{NJHigham2002KahanSum}.

\section{Initial conditions and simulation set} \label{secIII:InitCondSimSet}
\subsection{Initial conditions}
We compare five freely decaying incompressible flows, each initialized in the periodic box $[0,2\pi]^3$.
The cases are designed to span markedly different symmetries and large-scale organizations: a canonical nonhelical benchmark (TGV), helical structured flows (ABC-based), an ABC flow with an additional random-phase low-$k$ component, and a classical high-symmetry configuration known to generate intense small-scale activity (Kida--Pelz).
Unless otherwise noted, all initial fields have zero spatial mean, and the subsequent evolution is purely decaying (no forcing).

\HL{The Taylor--Green vortex is a canonical nonhelical flow with strong symmetries and is widely used as a transition-to-turbulence benchmark; its early evolution is characterized by the formation and roll-up of vortex sheets, followed by rapid generation of smaller scales. The Arnold--Beltrami--Childress (ABC) flow is a canonical helical large-scale flow. In its standard form it is a Beltrami flow, with vorticity aligned with velocity mode by mode, and it provides a convenient way to study how helical large-scale organization influences subsequent small-scale development. The Kida--Pelz configuration is a high-symmetry flow known for rapid gradient amplification and intense small-scale generation.}

\paragraph*{ABC building block.}
Our ABC-type initial conditions are constructed from the generalized Arnold--Beltrami--Childress (ABC) form
\begin{align}
\bfu_{\rm ABC}(x,y,z;\,A,B,C;\,k_x,k_y,k_z) \nonumber \\
= 
\begin{bmatrix}
A\sin(k_z z)+C\cos(k_y y) \\
B\sin(k_x x)+A\cos(k_z z) \\
C\sin(k_y y)+B\cos(k_x x)
\end{bmatrix},
\label{eq:IC_ABC}
\end{align}
which reduces to the classical ABC flow when $A=B=C=1$ and $k_x=k_y=k_z=1$ in a $2\pi$-periodic domain.
In all ABC-based cases below, we set $A=B=C=1$ for each ABC component and vary only the integer wavenumber triplets and superposition weights.

\paragraph*{(a) Multi-ABC.}
The multi-mode ABC case is the superposition of two ABC components,
\begin{eqnarray}
\bfu(x,y,z,0)
&=&
{w}_{1} \,\bfu_{\rm ABC}(x,y,z;\,1,1,1;\,1,1,1) \nonumber \\
&+&
{w}_{2} \,\bfu_{\rm ABC}(x,y,z;\,1,1,1;\,2,2,2),
\label{eq:IC_ABCMulti}
\end{eqnarray}
with weights $(w_1,w_2)=(1.0,0.5)$.
After the superposition, the field is rescaled to a target kinetic energy (see below) to facilitate consistent comparisons of amplitude-based diagnostics.

\paragraph*{(b) Multi-ABC (Asymmetric).}
To break the symmetry of the secondary ABC component, we replace the second wavenumber triplet in \eqref{eq:IC_ABCMulti} by an asymmetric choice,
\begin{eqnarray}
\bfu(x,y,z,0)
&=&
{w}_{1}\,\bfu_{\rm ABC}(x,y,z;\,1,1,1;\,1,1,1) \nonumber \\
&+&
{w}_{2}\,\bfu_{\rm ABC}(x,y,z;\,1,1,1;\,2,1,3),
\label{eq:IC_ABCMultiAsym}
\end{eqnarray}
again using $(w_1,w_2)=(1.0,0.5)$, followed by the same energy rescaling.

\paragraph*{(c) ABC + random-phase low-$k$.}
To test robustness against low-wavenumber randomness while retaining a structured large-scale backbone, we superpose an ABC field with a divergence-free random-phase perturbation supported only on low wavenumbers:
\begin{eqnarray}
\bfu(x,y,z,0)
&=&
w_{\rm abc}\,\bfu_{\rm ABC}(x,y,z;\,1,1,1;\,1,1,1) \nonumber \\
&+&
w_{\rm rnd}\,\bfu_{\rm rnd}(x,y,z;\,k_{\max}^{\rm rnd}),
\label{eq:IC_ABCplusRandom}
\end{eqnarray}
where $w_{\rm abc}=1.0$, $w_{\rm rnd}=0.3$, and $k_{\max}^{\rm rnd}=3$.
The random field $\bfu_{\rm rnd}$ is generated in Fourier space using random phases, projected to satisfy incompressibility mode-by-mode, and transformed back to physical space.
A fixed seed is used for reproducibility (seed $=42$).
As in the multi-ABC cases, the combined field is rescaled to the target kinetic energy.

\paragraph*{(d) Taylor--Green vortex (TGV).}
As a reference, we use the standard Taylor--Green initial condition \cite{TaylorGreen1937,Brachet1984,Pereira2021PRF}
\begin{equation}
\bfu(x,y,z,0)
=
\bigl(\sin x \cos y \cos z,\,
-\cos x \sin y \cos z,\,
0\bigr).
\label{eq:IC_TGV}
\end{equation}

\paragraph*{(e,f) Kida--Pelz.}
We use the standard high-symmetry Kida--Pelz initial condition \cite{Kida1985,Boratav1994},
\begin{align}
u_x(x,y,z,0) &= A_{\rm KP}\,\sin x \bigl(\cos 3y \cos z - \cos y \cos 3z\bigr), \nonumber \\
u_y(x,y,z,0) &= A_{\rm KP}\,\sin y \bigl(\cos 3z \cos x - \cos z \cos 3x\bigr), \nonumber \\
u_z(x,y,z,0) &= A_{\rm KP}\,\sin z \bigl(\cos 3x \cos y - \cos x \cos 3y\bigr),
\label{eq:IC_KidaPelz}
\end{align}
with a global amplitude factor $A_{\rm KP}$. In the present simulations we set $A_{\rm KP}=1.0$.
We perform two simulations for this initial condition: a base-resolution run at $512^3$ (case~(e)) and a higher-resolution run at $1024^3$ (case~(f)) used to assess the robustness of peak detection for the $k^4$-weighted diagnostic (details in Sec.~\ref{secIV:diagnostics}).

\subsection{Simulation parameters and data output}\label{subsec:simp_data_output}
The baseline set of simulations (cases (a)--(f)) is performed in the periodic domain $[0,2\pi]^3$ at viscosity $\nu=10^{-3}$. To probe viscosity dependence in a controlled setting, we additionally repeat the Taylor--Green vortex (TGV) initial condition at a lower viscosity $\nu=2.5\times 10^{-4}$ and at a higher viscosity $\nu=10^{-2}$ (see Sec.~\ref{subsec:results_viscosity}). The base resolution is $N^3=512^3$ for cases (a)--(e) and for the high-viscosity TGV run, while companion $N^3=1024^3$ runs are performed for Kida--Pelz (case~(f)) and for the low-viscosity TGV run in order to avoid cutoff-proximate peak locking in curvature-weighted spectra. Time integration uses a fixed time step $\Delta t=10^{-3}$.
Velocity (and, when needed, vorticity) snapshots are saved every $n_{\rm out}=50$ time steps for post-processing, corresponding to an output interval $\Delta t_{\rm out}=5\times 10^{-2}$.
The same spectral post-processing pipeline is applied to all cases (Sec.~\ref{secII:GovEqNumMET}), including the isotropic shell averaging conventions, the spherical two-thirds analysis mask, and transfer/flux diagnostics.

Because our focus is on timing relations and relative lead/lag measures, we choose the final integration time $t_{\rm end}$ in each case so that the dissipation episode is fully captured and the subsequent decay is sufficiently sampled.
Specifically, we integrate up to
$t_{\rm end}=60$ for the multi-ABC cases (a,b),
$t_{\rm end}=120$ for the ABC+random low-$k$ case (c),
and $t_{\rm end}=20$ for the baseline TGV and Kida--Pelz runs (d--f).
For the viscosity-variation TGV runs, we integrate up to $t_{\rm end}=16$ for $\nu=2.5\times 10^{-4}$
and up to $t_{\rm end}=12$ for $\nu=10^{-2}$.

For cross-case comparisons of amplitude-based quantities derived from spectra (e.g., $\varepsilon(t)$ and scale measures), we adopt the same calibration convention as in Ref.~\cite{TsuzukiTGVPrecursorPRF}: the spectra are rescaled such that the total kinetic energy at the first post-processed snapshot equals a reference value $K_0=1/8$.
This calibration does not affect peak \emph{wavenumbers} such as $k_{\rm peak}[\,|\curl\bfw|^2\,]$, and therefore does not alter the temporal ordering statements central to this work.

\begin{figure*}[t]
  \centering
  \includegraphics[width=\linewidth]{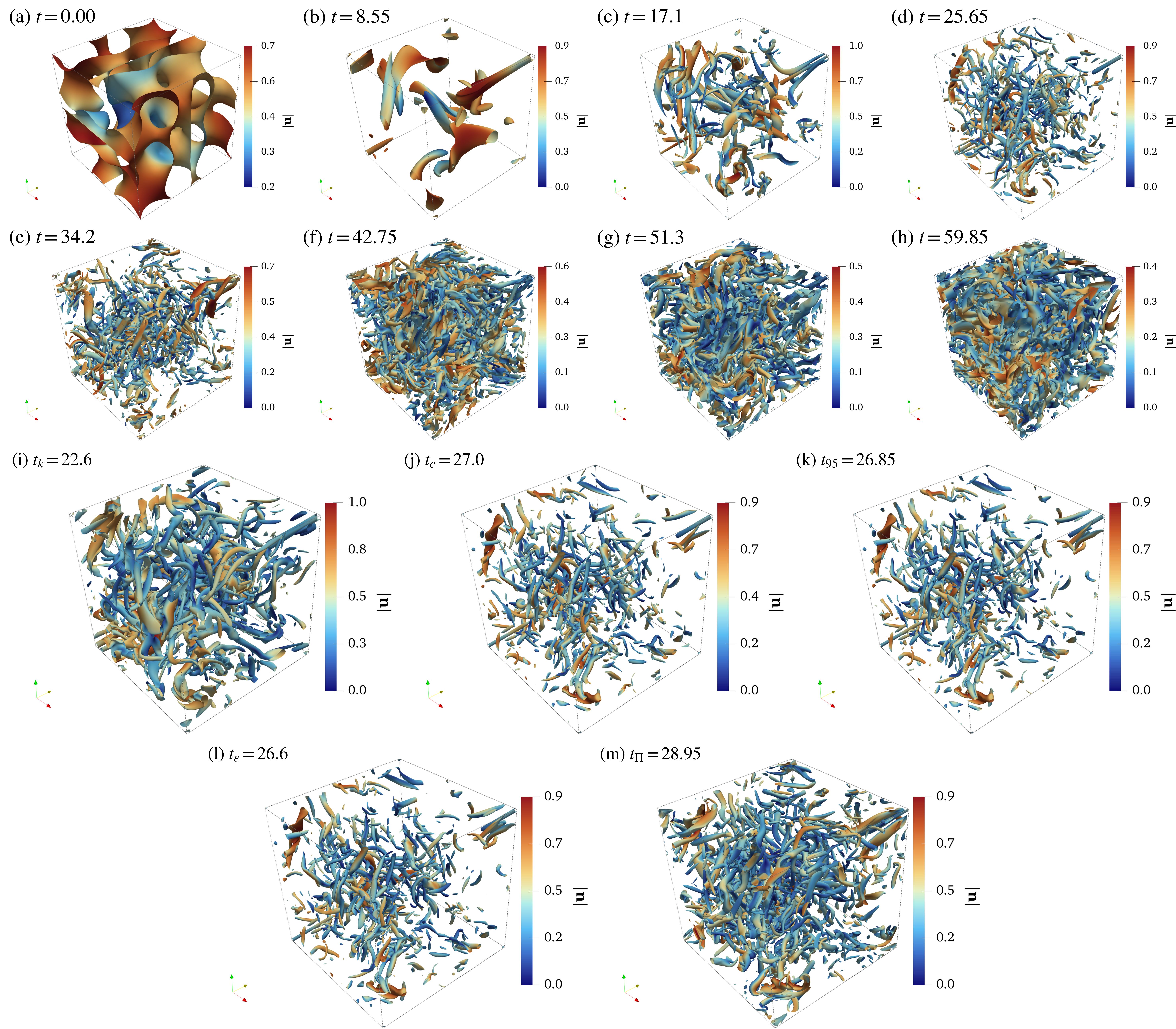}
  \caption{$Q$-criterion isosurfaces (colored by $|\bfu|$) at selected times for the $512^3$ Multi-ABC run.
Panels (a)--(h) show $t=0,\,8.55,\,17.1,\,25.65,\,34.2,\,42.75,\,51.3,\,59.85$.
Panels (i)--(m) show additional snapshots at characteristic times extracted from spectral diagnostics:
$t_k=22.6$, $t_c=27.0$, $t_{95}=26.85$, $t_\varepsilon=26.6$, and $t_\Pi=28.95$
(see Sec.~\ref{secIV:diagnostics} for definitions).
The isosurface level and rendering settings are the same as in Ref.~\cite{TsuzukiTGVPrecursorPRF}.}
  \label{fig:SimSnap_ABCMulti_512}
\end{figure*}

\begin{figure*}[t]
  \centering
  \includegraphics[width=\linewidth]{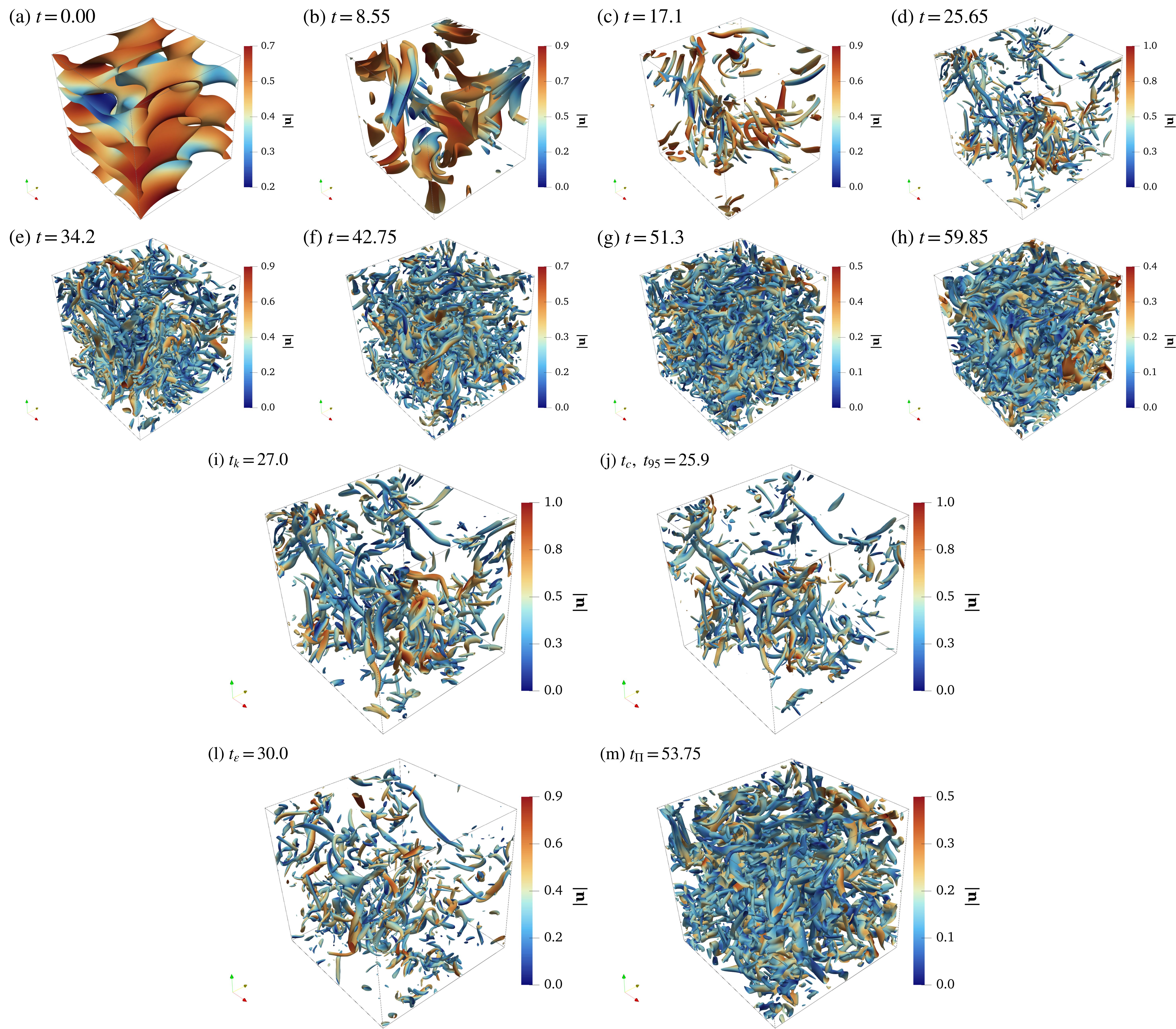}
\caption{$Q$-criterion isosurfaces (colored by $|\bfu|$) at selected times for the $512^3$ Multi-ABC (Asymmetric) run.
Panels (a)--(h) show $t=0,\,8.55,\,17.1,\,25.65,\,34.2,\,42.75,\,51.3,\,59.85$.
Panels (i)--(m) annotate snapshots at characteristic times:
$t_k=27.0$, $t_\varepsilon=30.0$, and $t_\Pi=53.75$ (Sec.~\ref{secIV:diagnostics}).
Here $t_c$ and $t_{95}$ coincide ($t_c=t_{95}=25.9$), and the corresponding snapshot is shown in panel~(j).
The isosurface level and rendering settings are the same as in Ref.~\cite{TsuzukiTGVPrecursorPRF}.}
\label{fig:SimSnap_ABCMultiAsym_512}
\end{figure*}

\begin{figure*}[t]
  \centering
  \includegraphics[width=\linewidth]{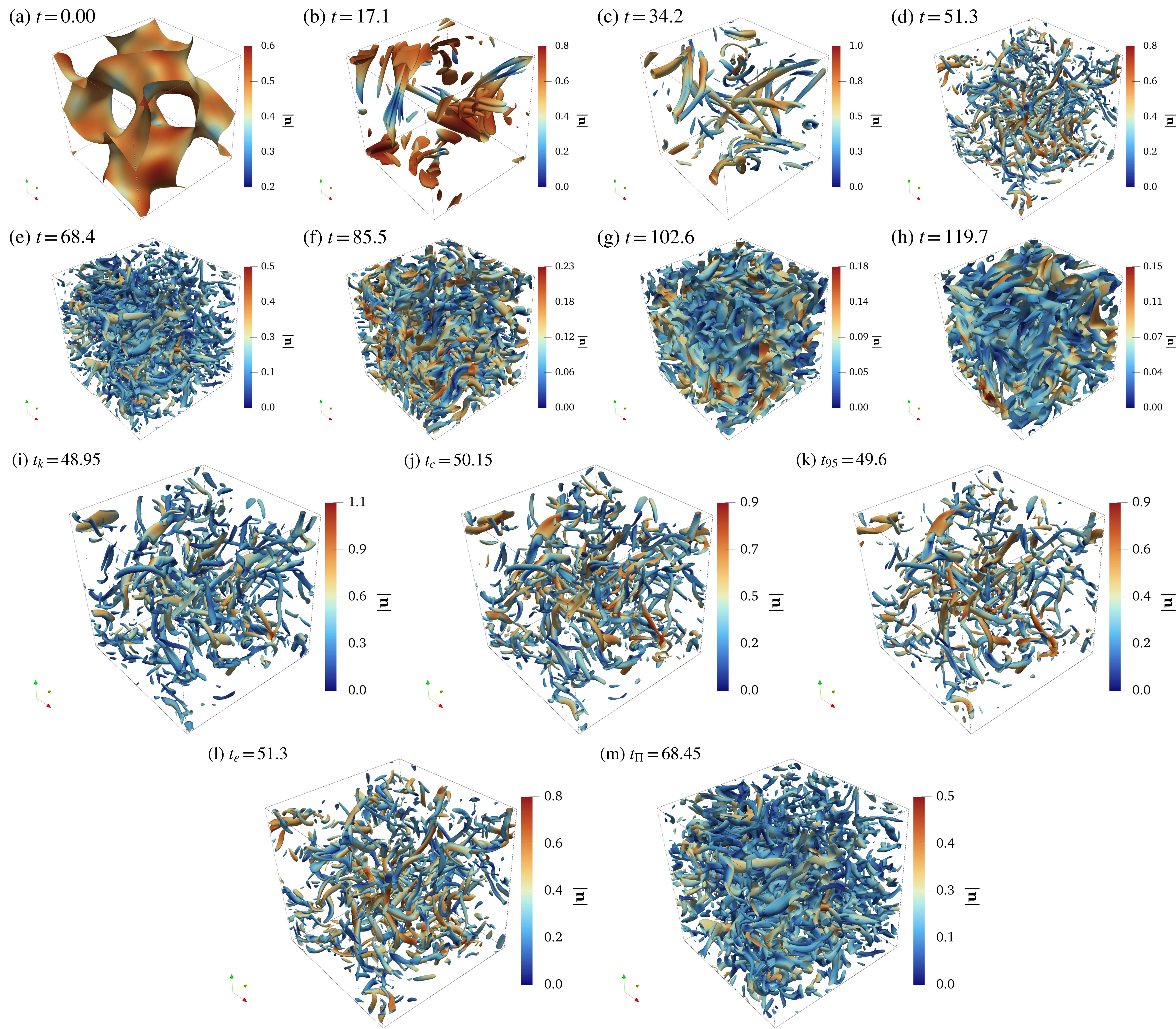}
  \caption{$Q$-criterion isosurfaces (colored by $|\bfu|$) at selected times for the $512^3$ ABC+random-phase low-$k$ run.
Panels (a)--(h) show $t=0,\,17.1,\,34.2,\,51.3,\,68.4,\,85.5,\,102.6,\,119.7$.
Panels (i)--(m) show snapshots at characteristic times:
$t_k=48.95$, $t_c=50.15$, $t_{95}=49.6$, $t_\varepsilon=51.3$, and $t_\Pi=68.45$
(see Sec.~\ref{secIV:diagnostics} for definitions).
The isosurface level and rendering settings are the same as in Ref.~\cite{TsuzukiTGVPrecursorPRF}.}
\label{fig:SimSnap_ABCRandomLowK_512}
\end{figure*}

\begin{figure*}[t]
  \centering
  \includegraphics[width=\linewidth]{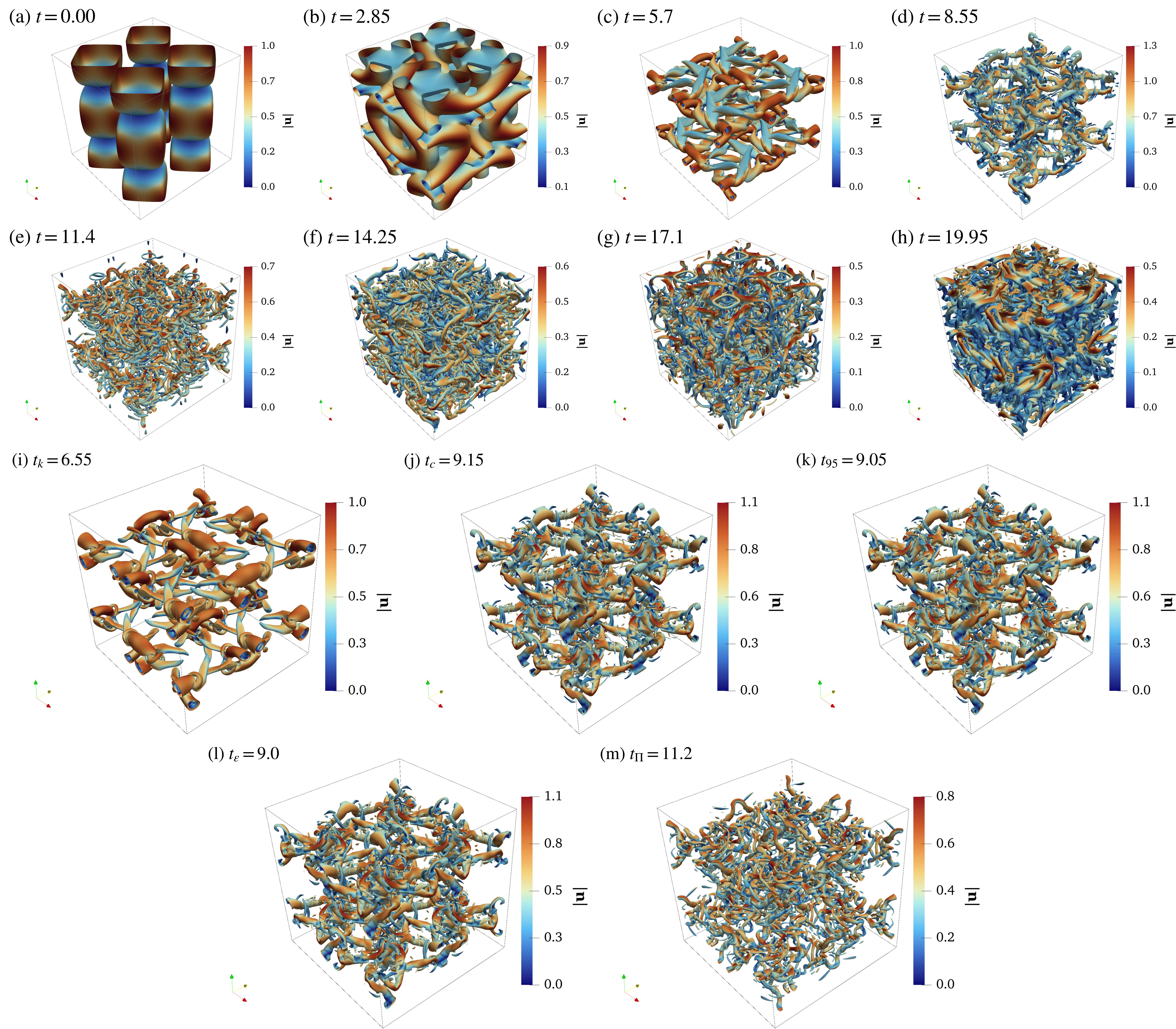}
  \caption{$Q$-criterion isosurfaces (colored by $|\bfu|$) at selected times for the $512^3$ Taylor--Green vortex (TGV) run.
Panels (a)--(h) show $t=0,\,2.85,\,5.7,\,8.55,\,11.4,\,14.25,\,17.1,\,19.95$.
Panels (i)--(m) show snapshots at characteristic times:
$t_k=6.55$, $t_c=9.15$, $t_{95}=9.05$, $t_\varepsilon=9.0$, and $t_\Pi=11.2$
(see Sec.~\ref{secIV:diagnostics} for definitions).
The isosurface level and rendering settings are the same as in Ref.~\cite{TsuzukiTGVPrecursorPRF}.}
\label{fig:SimSnap_TGV_512}
\end{figure*}

\begin{figure*}[t]
  \centering
  \includegraphics[width=\linewidth]{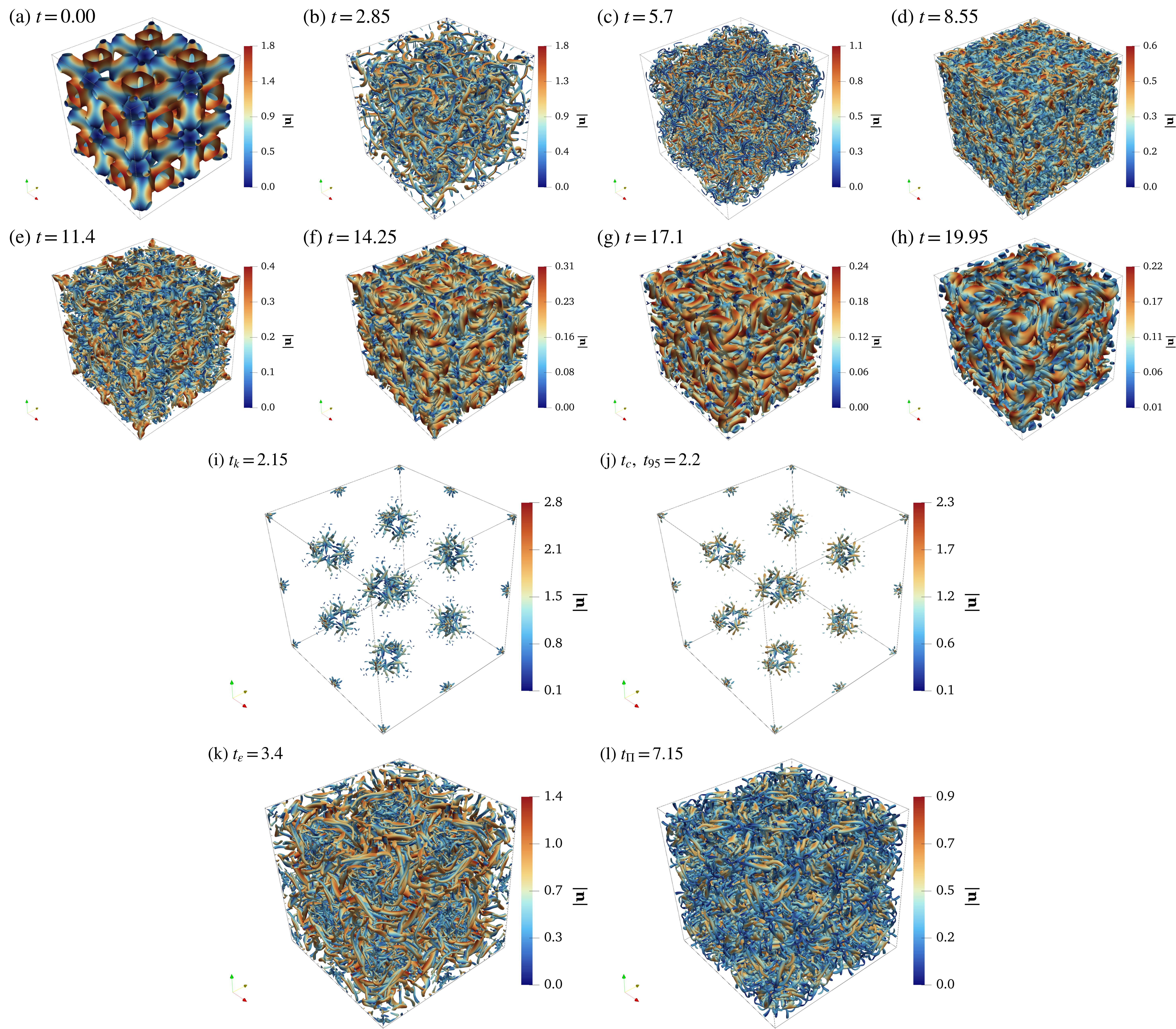}
\caption{$Q$-criterion isosurfaces (colored by $|\bfu|$) at selected times for the $512^3$ Kida--Pelz run.
Panels (a)--(h) show $t=0,\,2.85,\,5.7,\,8.55,\,11.4,\,14.25,\,17.1,\,19.95$.
Panels (i)--(l) show additional snapshots around the dissipation episode, including the dissipation- and flux-related times
$t_\varepsilon=3.4$ and $t_\Pi=7.15$ (Sec.~\ref{secIV:diagnostics}).
The $512^3$ Kida--Pelz run is shown here to highlight the rapid emergence of fine-scale structures and the resolution sensitivity of peak picking in curvature-weighted spectra; the characteristic times reported for Kida--Pelz in Table~\ref{tab:times_summary} are extracted from the $1024^3$ reference simulation (Sec.~\ref{subsec:spike-inspection}).
The isosurface level and rendering settings are the same as in Ref.~\cite{TsuzukiTGVPrecursorPRF}.
The snapshot labels (e.g., $t_k$, $t_c$, $t_{95}$) correspond to the $512^3$ visualization; $t_k$ differs from the $1024^3$ reference value due to peak-locking issues, whereas $t_{\varepsilon}$, $t_{c}$, $t_{95}$ were confirmed unchanged at $1024^{3}$ (Table~\ref{tab:aux_reaching_times}).
}
\label{fig:SimSnap_KidaPelz_512}
\end{figure*}

\paragraph*{Qualitative flow evolution in the $Q$-criterion visualizations.}
Figures~\ref{fig:SimSnap_ABCMulti_512}--\ref{fig:SimSnap_KidaPelz_512} show representative snapshots of the flow evolution for the five $512^3$ simulations.
In each panel, we plot an isosurface of the $Q$-criterion (with the same visualization settings as in Ref.~\cite{TsuzukiTGVPrecursorPRF}) and color it by the local speed $|\bfu|$.
The sequences in the upper rows (a)--(h) are chosen to span the transient from the initial large-scale organization to the subsequent volume-filling turbulent state and late-time decay (note that the physical time windows differ across cases, reflecting the different characteristic time scales of each initial condition).

For the multi-mode ABC case [Fig.~\ref{fig:SimSnap_ABCMulti_512}(a)--(h)], the initial field is dominated by smooth, domain-spanning structures characteristic of the superposed ABC modes.
As time increases, these structures rapidly fragment into twisted ribbons and then proliferate into a dense tangle of slender tube-like features that progressively occupies the full domain, indicating vigorous small-scale generation during the dissipation episode.
The asymmetric multi-mode ABC case [Fig.~\ref{fig:SimSnap_ABCMultiAsym_512}(a)--(h)] exhibits a qualitatively similar transition, while the broken symmetry yields a more heterogeneous spatial distribution and a less regular large-scale organization, with intermittent clustering of intense structures.

The ABC+random-phase low-$k$ case [Fig.~\ref{fig:SimSnap_ABCRandomLowK_512}(a)--(h)] displays a noticeably slower route to a volume-filling tangle.
Coherent, large-scale structures persist over a longer time interval, and the filamentary/tubular population becomes dense only at later times compared with the purely structured ABC-based cases, consistent with the delayed dissipation activity observed in the time-series diagnostics.

For the TGV [Fig.~\ref{fig:SimSnap_TGV_512}(a)--(h)], we recover the canonical progression from large-scale vortical sheets to rolled-up and interacting structures and finally to a densely tangled state, consistent with prior studies and our recent PRF Letter \cite{TsuzukiTGVPrecursorPRF}.
The Kida--Pelz case [Fig.~\ref{fig:SimSnap_KidaPelz_512}(a)--(h)] develops intense small-scale structures very rapidly.
Within a short time, the flow transitions from the highly symmetric initial arrangement to a domain-filling network of tightly curved tubes and sheets, underscoring the well-known propensity of this configuration to generate strong gradients \cite{Kida1985,Boratav1994}.

The lower panels (i) and onward in Figs.~\ref{fig:SimSnap_ABCMulti_512}--\ref{fig:SimSnap_KidaPelz_512} annotate additional snapshots at characteristic times extracted from the spectral diagnostics, including $t_k$ (the time at which the peak scale \HL{$k_{\mathrm{peak}}[\Cfour](t)$} attains its maximum), auxiliary reaching-time markers based on more global wavenumber measures (e.g., $t_c$ and $t_{95}$), as well as the dissipation- and flux-related times $t_\varepsilon$ and $t_\Pi$.
These times are defined precisely and analyzed quantitatively in Secs.~\ref{secIV:diagnostics} and~\ref{secV:results}, where we establish their consistent temporal ordering across initial conditions.
For the Kida--Pelz flow, the characteristic-time analysis reported below refers to the $1024^3$ simulation (Sec.~\ref{subsec:spike-inspection} and Table~\ref{tab:times_summary}); the $512^3$ visualization in Fig.~\ref{fig:SimSnap_KidaPelz_512} is retained to highlight the rapid emergence of fine-scale structures and the resolution sensitivity of peak picking in curvature-weighted spectra.

\section{Spectral diagnostics and characteristic times}\label{secIV:diagnostics}
\subsection{Isotropic spectra and calibration of the energy scale}
\label{subsec:spectra-calibration}

From the stored velocity snapshots, we construct isotropic (shell-summed) spectra in the periodic cube
$[0,2\pi]^3$.
As in our PRF Letter \cite{TsuzukiTGVPrecursorPRF}, shells are formed by binning Fourier modes by the integer
$k=\mathrm{round}(|\boldsymbol{\kappa}|)$ and accumulating plane-weighted sums appropriate for real-to-complex FFT layouts.
For the analysis, we additionally apply a \emph{spherical} two-thirds mask and record it as a binary column
\texttt{mask} in the spectra output (we restrict sums and peak searches to shells with \texttt{mask}$=1$ unless stated otherwise).
We denote the maximum retained shell by
\begin{equation}
k_{\max}^{(\mathrm{mask})} \equiv \max\{k:\texttt{mask}(k)=1\}, \label{eq:maskmax}
\end{equation}
which corresponds to the familiar spherical two-thirds cutoff (for $N=512$, $k_{\max}^{(\mathrm{mask})}=171$).

To enable a consistent comparison across different initial conditions, we use a constant calibration factor
$\alpha$ that rescales the spectra so that the total kinetic energy at the earliest stored time matches a target value
$K_0$:
\begin{equation}
\alpha \;=\; \frac{K_0}{\sum_k \E(k,t_0)} ,
\qquad
\E(k,t)\;\leftarrow\;\alpha\,\E(k,t).
\label{eq:alpha-calib}
\end{equation}
In the present study we take $K_0=1/8$, which coincides with the standard unit-amplitude TGV energy in $[0,2\pi]^3$.
All derived quantities below (including $\varepsilon$, $\eta$, and $R_\lambda$) are computed consistently from the calibrated spectra.

\subsection{Spectral quantities: $\E(k)$, $\D(k)$, \HL{$\PiC(q)$}, and \HL{$\Cfour(k)$}}
\label{subsec:spectral-quantities}

We track the standard spectral diagnostics of decaying turbulence:
the kinetic energy spectrum $\E(k,t)$, the dissipation density
\begin{equation}
\D(k,t)\;=\;2\nu k^2 \E(k,t),
\label{eq:Dk}
\end{equation}
and the dissipation rate
\begin{equation}
\varepsilon(t)\;=\;\sum_k \D(k,t)\;=\;2\nu\sum_k k^2 \E(k,t).
\label{eq:eps}
\end{equation}
We also analyze the nonlinear transfer spectrum $T(k,t)$ and the cumulative flux
\HL{
\begin{equation}
\PiC(q,t)\;=\;-\sum_{m\le q} T(m,t),
\label{eq:Pi-def}
\end{equation}
}
so that \HL{$|\PiC(q,t)|$} measures the magnitude of interscale transfer across the \HL{cutoff wavenumber~$q$}.

The central object of this work is the \HL{shell-summed curl-of-vorticity spectrum $\Cfour(k,t)$, defined in Eq.~\eqref{eq:curlw2_equiv_methods} with $\bfw \equiv \curl \bfu$}.
For incompressible flow, $\Div\bfu=0$ implies \HL{$\curl\bfw=-\nabla^2\bfu$, so $\Cfour(k,t)$ corresponds to the modal $|\boldsymbol{\kappa}|^4|\hat{\bfu}|^2$ weight and is denoted by the shorthand}
\HL{
\begin{equation}
\Cfour(k,t)\simeq k^4\E(k,t)
\label{eq:curlw2-k4E}
\end{equation}
}
\HL{when expressed in terms of the integer shell center}.
This $k^4$ weighting accentuates the incipient formation of high-curvature structures (tightly wound tubes and rolled-up sheets), which, as we show below, advances and saturates in wavenumber \emph{before} $\varepsilon(t)$ reaches its maximum.

\subsection{Integral and small-scale measures: $L_{\mathrm{int}}$, $\lambda$, $\eta$, $R_\lambda$, and $k_{\max}\eta$}
\label{subsec:scales-resolution}

We monitor four complementary measures of the evolving length scales.
First, the instantaneous Kolmogorov scale is
\begin{equation}
\eta(t)\;=\;\big(\nu^3/\varepsilon(t)\big)^{1/4}.
\label{eq:eta}
\end{equation}
Second, the integral scale is evaluated from the isotropic spectrum as
\begin{equation}
L_{\mathrm{int}}(t)\;=\;\frac{\pi}{2\sum_k \E(k,t)}\sum_{k\ge 1}\frac{\E(k,t)}{k},
\label{eq:Lint}
\end{equation}
which emphasizes the energy-containing range.
Third, the Taylor microscale is computed as
\begin{equation}
\lambda(t)\;=\;\left(\frac{\sum_k \E(k,t)}{\sum_k k^2 \E(k,t)}\right)^{1/2},
\label{eq:lambda}
\end{equation}
and with $u'(t)=\sqrt{\tfrac{2}{3}\sum_k \E(k,t)}$ we define the Taylor-scale Reynolds number
\begin{equation}
R_\lambda(t)\;=\;\frac{u'(t)\,\lambda(t)}{\nu}.
\label{eq:Rlambda}
\end{equation}
\HL{To quantify the non-dimensional dissipation level in this unsteady decay, we also report}
\HL{
\begin{equation}
C_{\varepsilon}(t)
\;=\;
\frac{\varepsilon(t)L_{\mathrm{int}}(t)}{u'(t)^3}.
\label{eq:Ceps}
\end{equation}
}
\HL{This quantity is useful in unsteady turbulence because it indicates how far the decay departs from a quasi-equilibrium picture in which a nearly constant normalized dissipation might be expected.}
Finally, to quantify resolution, we track the standard criterion $k_{\max}\eta(t)$, where we follow Ref.~\cite{TsuzukiTGVPrecursorPRF}
and define the maximum wavenumber by the Nyquist value $k_{\max}=(N/2)\,2\pi/L$ (thus $k_{\max}=N/2$ in our $L=2\pi$ domain).
In Fig.~\ref{fig:kmaxeta-diagnostics} we plot $R_\lambda(t)$ together with $k_{\max}\eta(t)$.
All cases satisfy $k_{\max}\eta(t)\gtrsim \mathcal{O}(1)$ throughout the time intervals analyzed, and the minimum typically occurs near the dissipation peak.
Figure~\ref{fig:scales-variations} summarizes the concurrent evolution of $L_{\mathrm{int}}(t)$, $\lambda(t)$, $\eta(t)$, $\varepsilon(t)$, \HL{and $C_{\varepsilon}(t)$,}
highlighting that $\eta(t)$ reaches its minimum at $t=t_\varepsilon$ while the large-scale measures evolve more gradually.
\HL{Representative velocity spectra at selected times are shown in Appendix Fig.~\ref{fig:appendix-velocity-spectra}.
For each case, we plot $E(k,t)$ at $t=0$, $t_k$, $t_\varepsilon$, $t_\Pi$, and a late-time snapshot.
These spectra show how the initially large-scale energy distribution broadens toward higher wavenumbers during the transient decay and provide a spectral complement to the scale and dissipation histories in Fig.~\ref{fig:scales-variations}.}

\subsection{$k$-metrics for spectral ``peak scales'': peak, centroid, and quantiles}
\label{subsec:k-metrics}

To quantify how far toward small scales a given diagnostic has progressed, we introduce several $k$-metrics
for a nonnegative shell spectrum $S(k,t)$.
The most direct measure is the peak wavenumber
\begin{equation}
k_{\mathrm{peak}}[S](t)\;\equiv\;\arg\max_{k\ge k_{\min}} S(k,t),
\label{eq:kpeak}
\end{equation}
where we take $k_{\min}=1$.
We also use the centroid scale
\begin{equation}
k_{\mathrm{centroid}}[S](t)\;\equiv\;\frac{\sum_k k\,S(k,t)}{\sum_k S(k,t)},
\label{eq:kcentroid}
\end{equation}
and the weighted quantile scale $k_q[S](t)$ defined by the cumulative weight:
\begin{equation}
k_q[S](t)\;\equiv\;\min\left\{k:\frac{\sum_{m\le k} S(m,t)}{\sum_m S(m,t)}\ge q\right\},
\qquad 0<q<1.
\label{eq:kq}
\end{equation}
In particular, we will refer frequently to $k_{95}[S](t)$.

In the present work, we apply these definitions to two spectra:
\begin{enumerate}
\item \HL{$S(k,t)=\Cfour(k,t)$ (the curl-of-vorticity/palinstrophy-related} spectrum), and
\item $S(k,t)=|\PiC(k,t)|$ (the magnitude of the cumulative flux).
\end{enumerate}
Figures~\ref{fig:peaks-variations-ABCMulti}--\ref{fig:peaks-variations-KidaPelzHighRes} show the time evolution of the corresponding $k$-metrics for all cases: (a) Multi-ABC, (b) Multi-ABC (Asymmetric), (c) ABC + random-phase low-$k$, (d) TGV, (e) Kida--Pelz, and (f) Kida--Pelz [1024$^{3}$], together with $\varepsilon(t)$.
Across initial conditions, the different definitions (peak/centroid/quantile) yield consistent qualitative trends,
with the peak metric providing the most direct ``advance-and-level-off'' indicator.

\subsection{Characteristic times: $t_\varepsilon$, $t_k$, and $t_\Pi$} \label{subsec:characteristic-times}

We define the dissipation peak time as
\begin{equation}
t_\varepsilon \;\equiv\; \arg\max_{t}\, \varepsilon(t).
\end{equation}
For the peak-based diagnostics, we define the characteristic times by the maximum-attainment times of the peak scales:
\HL{
\begin{align}
t_k &\;\equiv\; \arg\max_{t}\, k_{\mathrm{peak}}[\Cfour](t), \\
t_\Pi &\;\equiv\; \arg\max_{t}\, k_{\mathrm{peak}}[|\PiC|](t).
\end{align}
}
When the maximum is attained over multiple consecutive outputs (a plateau), we take the earliest occurrence.
For the TGV this is consistent with the ``advance-and-level-off'' onset time used in our Letter, because the rise of
\HL{$k_{\mathrm{peak}}[\Cfour](t)$} terminates in a short plateau near its maximum.

For visualization and as auxiliary reference values in the $Q$-snapshot figures (Figs.~\ref{fig:SimSnap_ABCMulti_512}--\ref{fig:SimSnap_KidaPelz_512}),
we also report reaching times for more global metrics such as $k_{\mathrm{centroid}}$ and $k_{95}$.
For a generic $k$-metric $k_m(t)$, we define $k_m^\varepsilon \equiv k_m(t_\varepsilon)$ and, for a threshold $q\in(0,1)$,
\begin{equation}
t_{m,q} \;\equiv\;
\min\left\{t:\;\; k_m(t') \ge q\,k_m^\varepsilon\;\; \forall t'\in[t,t+M\Delta t_{out}]\right\},
\label{eq:reaching-time}
\end{equation}
where $\Delta t_{out}$ is the output interval and we take $M=3$.
In the snapshots we denote \HL{$t_c \equiv t_{\,k_{\mathrm{centroid}}[\Cfour],\,q}$ and
$t_{95} \equiv t_{\,k_{95}[\Cfour],\,q}$} with $q=0.9$.
Because the reaching-time criterion depends on the choice of $(q,M)$, we use $t_c$ and $t_{95}$ only as auxiliary markers and do not rely on them for the primary ordering arguments.

All characteristic times reported in this work are extracted from time series sampled at the output cadence 
$\Delta t_{\rm out}=5\times10^{-2}$ (Sec.~\ref{subsec:simp_data_output}). Therefore, $t_k$, $t_\varepsilon$, and $t_\Pi$ are quantized in steps of
$\Delta t_{\rm out}$ and carry an intrinsic temporal discretization uncertainty of ${\cal O}(\Delta t_{\rm out})$.
Conservatively, we take $|\delta t_k|,|\delta t_\varepsilon|,|\delta t_\Pi|\le \Delta t_{\rm out}$, which implies
$|\delta\Delta t_{k\varepsilon}|,|\delta\Delta t_{\varepsilon\Pi}|\le 2\Delta t_{\rm out}$ for the lead/lag measures.
In all cases summarized in Table~\ref{tab:times_summary}, the observed separations exceed this bound; in particular, the smallest lead time
$\Delta t_{k\varepsilon}=0.40$ corresponds to $8\,\Delta t_{\rm out}$, so the ordering is robust to the output cadence.

\HL{The peak-based characteristic times are extracted from the raw, unsmoothed shell spectra. We use the raw spectra because smoothing would introduce an additional parameter and could mask cutoff-proximate peak locking, which is itself an important resolution diagnostic for high-wavenumber-weighted spectra. When the maximum is attained over multiple consecutive output times, we take the earliest occurrence, consistent with the interpretation of $t_k$ as the end of the initial advance of the peak scale.}

In all cases of Figs.~\ref{fig:peaks-variations-ABCMulti}--\ref{fig:peaks-variations-KidaPelzHighRes}, $t_k$ consistently precedes $t_\varepsilon$, while $t_\Pi$ occurs later than $t_\varepsilon$,
establishing the ordering $t_k<t_\varepsilon<t_\Pi$ across the tested initial conditions.
The centroid- and quantile-type measures ($k_{\mathrm{centroid}}$ and $k_{95}$) exhibit the same qualitative behavior, providing an additional robustness check beyond a single peak definition.

\subsection{Inspection for high-$k$ locking and cutoff effects}
\label{subsec:spike-inspection}

Because \HL{$\Cfour(k,t)$} involves a strong $k^4$ weighting, peak-based indicators can be
particularly sensitive to finite-resolution effects in the high-$k$ range.
In extreme transient states, this sensitivity may lead to spurious peak detection
if the argmax is influenced by spectral content close to the analysis cutoff.
To assess this possibility, we explicitly inspect \HL{$\Cfour(k,t)$} at representative times
and mark the detected peak locations, as summarized in Fig.~\ref{fig:inspections}.

For the $512^3$ ABC-based cases and the TGV at the baseline viscosity $\nu=10^{-3}$, the spectrum exhibits a clear maximum of
\HL{$\Cfour(k,t)$} that is well separated from the cutoff.
In contrast, for the Kida--Pelz case at $512^3$, the instantaneous argmax can approach
$k_{\max}^{(\mathrm{mask})}$ during the early transient, indicating a cutoff-proximate
``locking'' of the peak.
This behavior is visualized explicitly in Fig.~\ref{fig:inspections}(e) and reflects
the combination of intense small-scale generation in the Kida--Pelz flow and the strong
high-$k$ bias inherent in the $k^4$ weighting.

In this work we treat cutoff-proximate peaks as a \emph{resolution warning} rather than as
a feature to be ``corrected'' by an alternative peak definition.
This inspection protocol is applied throughout the paper to all simulations reported here, including the viscosity-variation TGV runs in Sec.~\ref{subsec:results_viscosity}. When cutoff-proximate locking is encountered (as in the $512^3$ Kida--Pelz case and the low-viscosity TGV case), we do not attempt to suppress it by an ad hoc restriction of the peak-search interval; instead, we treat it as a resolution warning and rely on a higher-resolution companion simulation for quantitative comparison.

Accordingly, to verify that the near-cutoff peak locking observed in the $512^3$ Kida--Pelz run is not physical, we performed a companion simulation at $1024^3$.
As shown in Fig.~\ref{fig:inspections}(f), the higher-resolution spectrum exhibits a well-defined peak away from the cutoff and shows no tendency toward high-$k$ locking.
This confirms that the issue observed at $512^3$ originates from limited resolution rather than from
the underlying dynamics.

Based on this assessment, the Kida--Pelz results discussed in the remainder of this paper
are taken from the $1024^3$ simulation, while the $512^3$ case is retained solely to
demonstrate the resolution sensitivity of peak detection in curvature-weighted spectra.

\section{Results}\label{secV:results}
\subsection{Resolution diagnostics and Reynolds-number evolution}
\label{subsec:results_resolution}

\begin{figure*}[t]
  \centering
  \includegraphics[width=\linewidth]{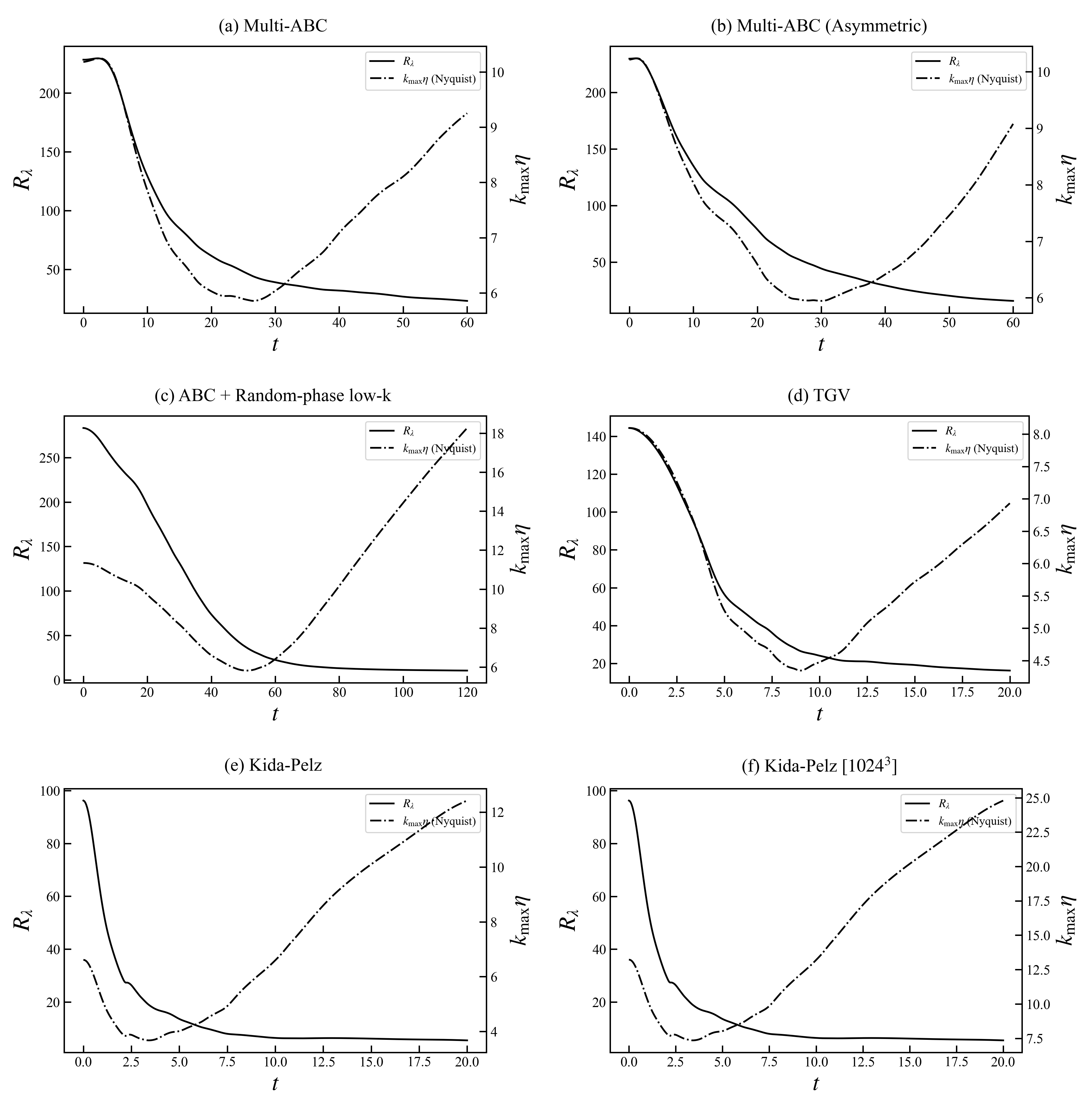}
  \caption{Resolution diagnostics for the six simulations:
time series of the Taylor-microscale Reynolds number $R_\lambda(t)$ (solid line, left axis)
and the resolution measure $k_{\max}\eta(t)$ (dash-dotted line, right axis; Nyquist definition $k_{\max}=(N/2)\,2\pi/L$).
Panels correspond to (a) Multi-ABC, (b) Multi-ABC (Asymmetric), (c) ABC+random-phase low-$k$,
(d) TGV, (e) Kida--Pelz ($512^3$), and (f) Kida--Pelz ($1024^3$).
The Kolmogorov length $\eta(t)=(\nu^3/\varepsilon)^{1/4}$ and $R_\lambda(t)$ are computed from the isotropic spectra as described in Sec.~\ref{secIV:diagnostics}.}
\label{fig:kmaxeta-diagnostics}
\end{figure*}

We begin by confirming that all runs are adequately resolved over the time intervals analyzed.
Figure~\ref{fig:kmaxeta-diagnostics} shows the time series of the Taylor-scale Reynolds number
$R_\lambda(t)$ together with the resolution indicator $k_{\max}\eta(t)$, where $k_{\max}=(N/2)\,2\pi/L$ is the Nyquist wavenumber (as in Ref.~\cite{TsuzukiTGVPrecursorPRF})
and $\eta(t)=(\nu^3/\varepsilon(t))^{1/4}$ is the Kolmogorov length.
For all $512^3$ cases (Multi-ABC, Multi-ABC asymmetric, ABC+random-phase low-$k$, TGV, and Kida--Pelz),
$k_{\max}\eta(t)$ remains above unity throughout the evolution, including around the dissipation episode,
indicating that the smallest dynamically relevant scales are captured.
The concomitant decrease of $R_\lambda(t)$ during the dissipation surge reflects the rapid development of
small scales and the accompanying increase of the characteristic gradients.
\HL{The increase of $R_\lambda$ at early times in some cases should not be interpreted as evidence of continuing external production. After initialization, all runs evolve without forcing and satisfy a freely decaying energy budget. The early-time rise of $R_\lambda$ instead reflects a transient cascade-development stage in which $u'$, $\lambda$, and the small-scale gradients evolve at different rates while the flow reorganizes toward its strongest dissipative state.}

For the Kida--Pelz initial condition, we additionally performed a $1024^3$ simulation (shown as the
high-resolution counterpart in Fig.~\ref{fig:kmaxeta-diagnostics}) to address a peak-detection issue in
the curvature-weighted spectrum (discussed below).
As expected, the higher-resolution run provides a substantially larger $k_{\max}\eta(t)$ margin
over the entire time interval, which helps disentangle physical peak motion from finite-resolution artifacts.

\subsection{Evolution of integral and dissipative scales}
\label{subsec:results_scales}

Figure~\ref{fig:scales-variations} summarizes the evolution of representative scale measures
$L_{\mathrm{int}}(t)$, the Taylor microscale $\lambda(t)$, the Kolmogorov scale $\eta(t)$, the
dissipation rate $\varepsilon(t)$\HL{, and the normalized dissipation coefficient $C_{\varepsilon}(t)$} for all cases.
\HL{The additional $C_{\varepsilon}(t)$ row is included because the present flows are unsteady and can depart from classical quasi-equilibrium dissipation phenomenology.}
\HL{The plotted $C_{\varepsilon}(t)$ histories are not approximately flat: in most cases $C_{\varepsilon}(t)$ changes markedly through the approach to $t_\varepsilon$ and continues to evolve during the subsequent decay. This non-constant behavior supports the non-equilibrium interpretation of the present decays and motivates treating dissipation-peak timing as a transient diagnostic rather than as a quasi-steady estimate.}
Across initial conditions, the same qualitative pattern is observed: as the flow transitions from
its initially organized large-scale state toward a turbulent, volume-filling state, the energy-containing
scale $L_{\mathrm{int}}$ and the intermediate scale $\lambda$ decrease, while $\eta$ decreases toward
its minimum near the dissipation peak and then increases during late-time decay.

The principal difference across cases is the characteristic time window over which this transient unfolds.
The structured ABC-based initial conditions (Multi-ABC and its asymmetric variant) exhibit a comparatively
rapid onset of small-scale activity followed by a long decay, whereas the ABC+random-phase low-$k$ case
develops and dissipates more slowly, consistent with the prolonged persistence of large-scale organization
seen in the $Q$-criterion snapshots (Sec.~\ref{secIII:InitCondSimSet}).
The TGV evolution reproduces the canonical timescale separation reported previously \cite{TsuzukiTGVPrecursorPRF},
while the Kida--Pelz flow produces strong gradients very rapidly, in line with the known propensity of
this highly symmetric configuration to generate intense small-scale structures \cite{Kida1985,Boratav1994}.

These scale trends provide an important backdrop for interpreting curvature-weighted spectral indicators:
the precursor times identified below consistently occur during the phase in which $\eta(t)$ is still
decreasing and the flow morphology is transitioning toward filament/tube-dominated structures.

\HL{Appendix Fig.~\ref{fig:appendix-velocity-spectra} complements Fig.~\ref{fig:scales-variations} by showing representative velocity energy spectra $E(k,t)$ at $t=0$, $t_k$, $t_\varepsilon$, $t_\Pi$, and a late-time snapshot for each case. These spectra show directly how kinetic energy broadens from the initial large-scale distribution toward higher wavenumbers during the transient decay.}

\subsection{Spike inspection for the curvature-weighted spectrum}
\label{subsec:results_inspection}

A central observable in this work is the curvature-weighted spectrum
\HL{$\Cfour(k,t)$} (equivalently $k^4\E(k)$ for incompressible flow), and the associated peak/centroid/quantile
wavenumber measures introduced in Sec.~\ref{secIV:diagnostics}.
Because \HL{$\Cfour(k,t)$} strongly emphasizes high wavenumbers, its instantaneous argmax can be particularly sensitive to
finite-resolution effects near the analysis cutoff $k_{\max}^{(\mathrm{mask})}$.
Following the inspection protocol described in Sec.~\ref{subsec:spike-inspection}, we therefore perform an explicit
\emph{spike inspection}: we examine representative spectra and verify whether the detected maximizer lies well inside the
resolved range. If the maximizer lies at (or very close to) $k_{\max}^{(\mathrm{mask})}$, we interpret this as a
\emph{resolution warning} and rely on a higher-resolution companion simulation for quantitative extraction of
characteristic times. Importantly, we do \emph{not} remove, trim, or otherwise exclude peaks; the inspection is used
only to flag cutoff-proximate peak locking.

\begin{figure*}[t]
  \centering
  \includegraphics[width=\linewidth]{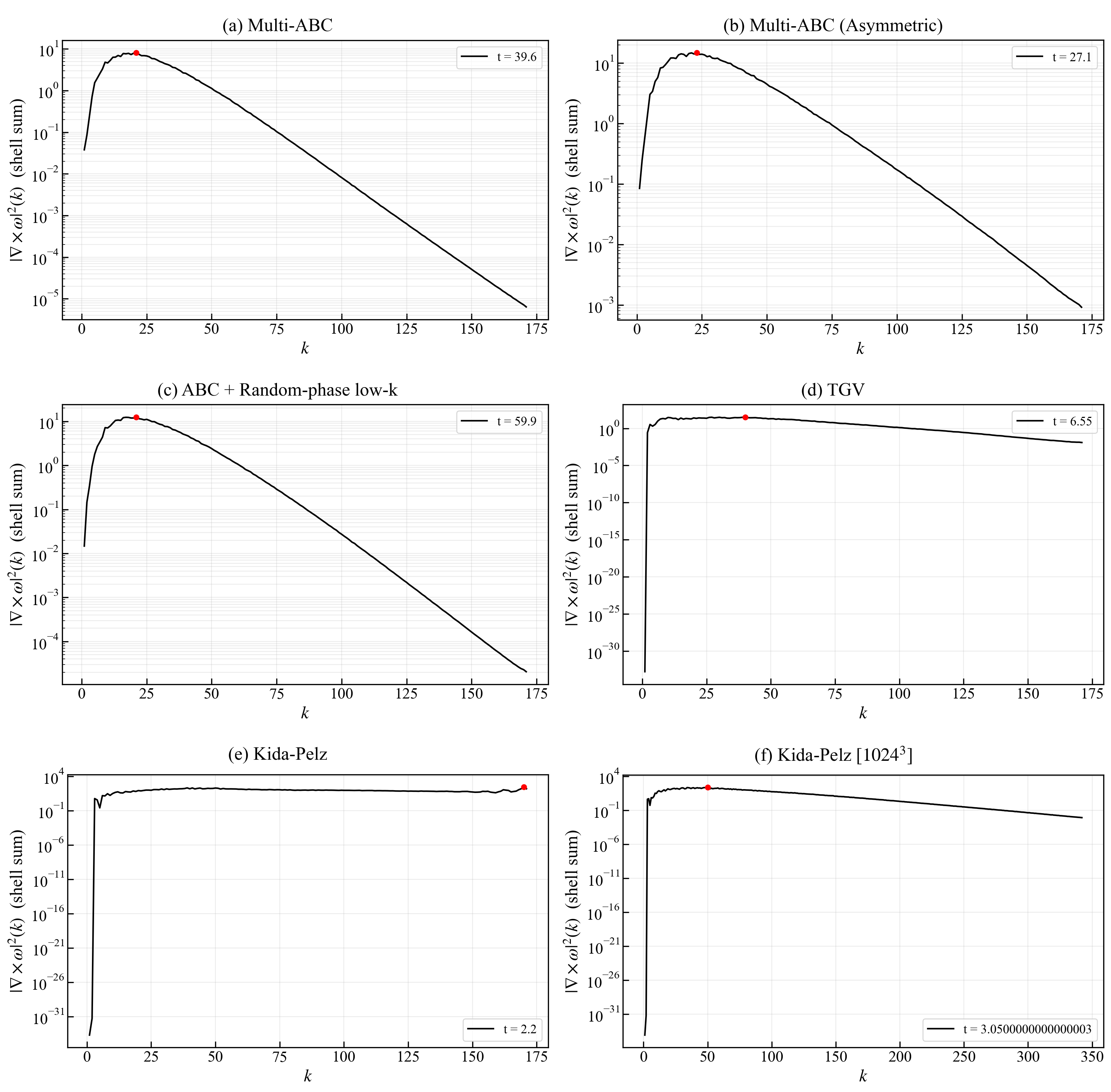}
  \caption{Inspection of peak-scale detection based on the shell-summed \HL{curl-of-vorticity spectrum $\Cfour(k,t)$}. Red markers denote \HL{the detected maximizer $k_{\mathrm{peak}}[\Cfour](t)$. The panels are intended to show whether the detected peak lies well inside the retained spectral range or is influenced by the high-wavenumber cutoff}. The $512^3$ Kida--Pelz case exhibits \HL{cutoff-proximate peak locking, whereas} the \HL{$1024^3$ Kida--Pelz reference run shows} a \HL{well-defined peak away from} the cutoff\HL{. Thus Fig}.~\HL{\ref{fig:inspections} is} used as \HL{a resolution-inspection diagnostic rather than as a comparison of} the \HL{far spectral tail over many decades}.}
\label{fig:inspections}
\end{figure*}

Figure~\ref{fig:inspections} presents representative inspections of \HL{$\Cfour(k,t)$} at selected times, with the detected
peak marked.
For the Multi-ABC, Multi-ABC (Asymmetric), ABC+random-phase low-$k$, and TGV cases, the spectrum exhibits a well-defined
maximum at intermediate-to-small scales at the inspected time, and the peak is clearly separated from
$k_{\max}^{(\mathrm{mask})}$ (the spherical two-thirds cutoff).
In contrast, for the Kida--Pelz case at $512^3$, the maximizer can approach $k_{\max}^{(\mathrm{mask})}$ during the early
transient, indicating cutoff-proximate peak locking [Fig.~\ref{fig:inspections}(e)].
This endpoint hit is also visible in the time-series view in Fig.~\ref{fig:peaks-variations-KidaPelz}, where
\HL{$k_{\mathrm{peak}}[\Cfour](t)$} can reach $k_{\max}^{(\mathrm{mask})}$.
Accordingly, we use the $1024^3$ Kida--Pelz data (Figs.~\ref{fig:inspections}(f) and~\ref{fig:peaks-variations-KidaPelzHighRes}) as the
reference for the Kida--Pelz precursor analysis, while retaining the $512^3$ case only to illustrate this finite-resolution
failure mode.

\subsection{Precursor behavior and robust temporal ordering across initial conditions}
\label{subsec:results_ordering}
\begin{figure*}[t]
  \centering
  \includegraphics[width=\linewidth]{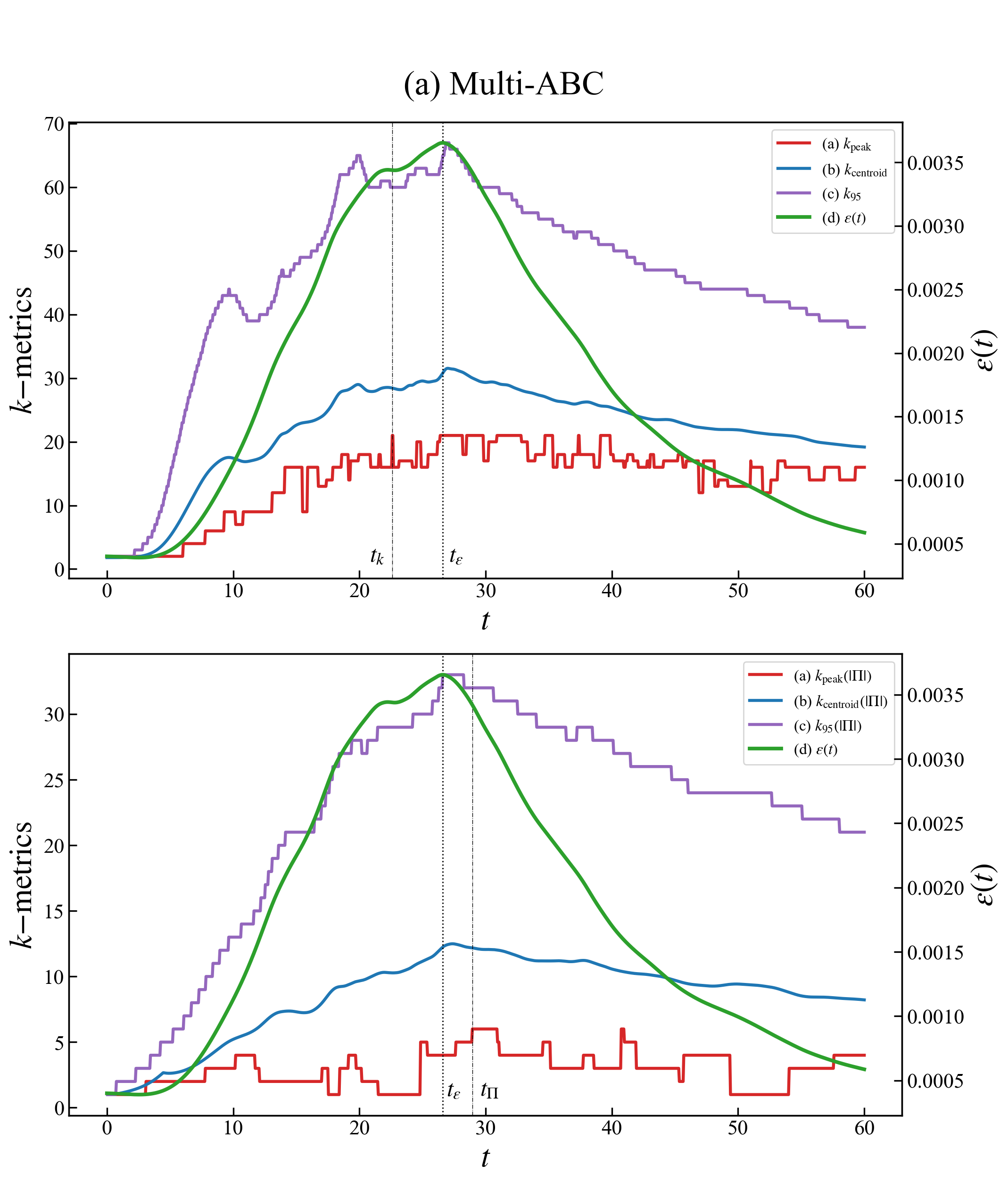}
  \caption{\textbf{Summary of peak/centroid/quantile wavenumber measures and their timing in (a) Multi-ABC.}
The \emph{upper} panel shows characteristic wavenumber measures extracted from the curvature-weighted spectrum
\HL{$\Cfour(k,t)$} (equivalently $k^4\E(k)$): $k_{\mathrm{peak}}$ (red), $k_{\mathrm{centroid}}$ (blue), and $k_{95}$ (purple),
together with the dissipation rate $\varepsilon(t)$ (green, right axis).
The \emph{lower} panel shows the corresponding measures computed from the flux diagnostic \HL{$|\PiC(q)|$}
(denoted $k_{\mathrm{peak}}(\Pi)$, $k_{\mathrm{centroid}}(\Pi)$, and $k_{95}(\Pi)$), again with $\varepsilon(t)$ for reference.
Vertical markers indicate the characteristic times defined in Sec.~\ref{secIV:diagnostics}: \HL{$t_k=\arg\max_t\,k_{\mathrm{peak}}[\Cfour](t)$}, $t_\varepsilon=\arg\max_t\,\varepsilon(t)$, and $t_\Pi=\arg\max_t\,k_{\mathrm{peak}}[|\PiC|](t)$.
This combined view enables a direct comparison between (i) the early progression of curvature-weighted small-scale signatures and
(ii) the later evolution of interscale transfer, demonstrating the consistent temporal ordering $t_k<t_\varepsilon<t_\Pi$ across all initial conditions.}
\label{fig:peaks-variations-ABCMulti}
\end{figure*}

\begin{figure*}[t]
  \centering
  \includegraphics[width=\linewidth]{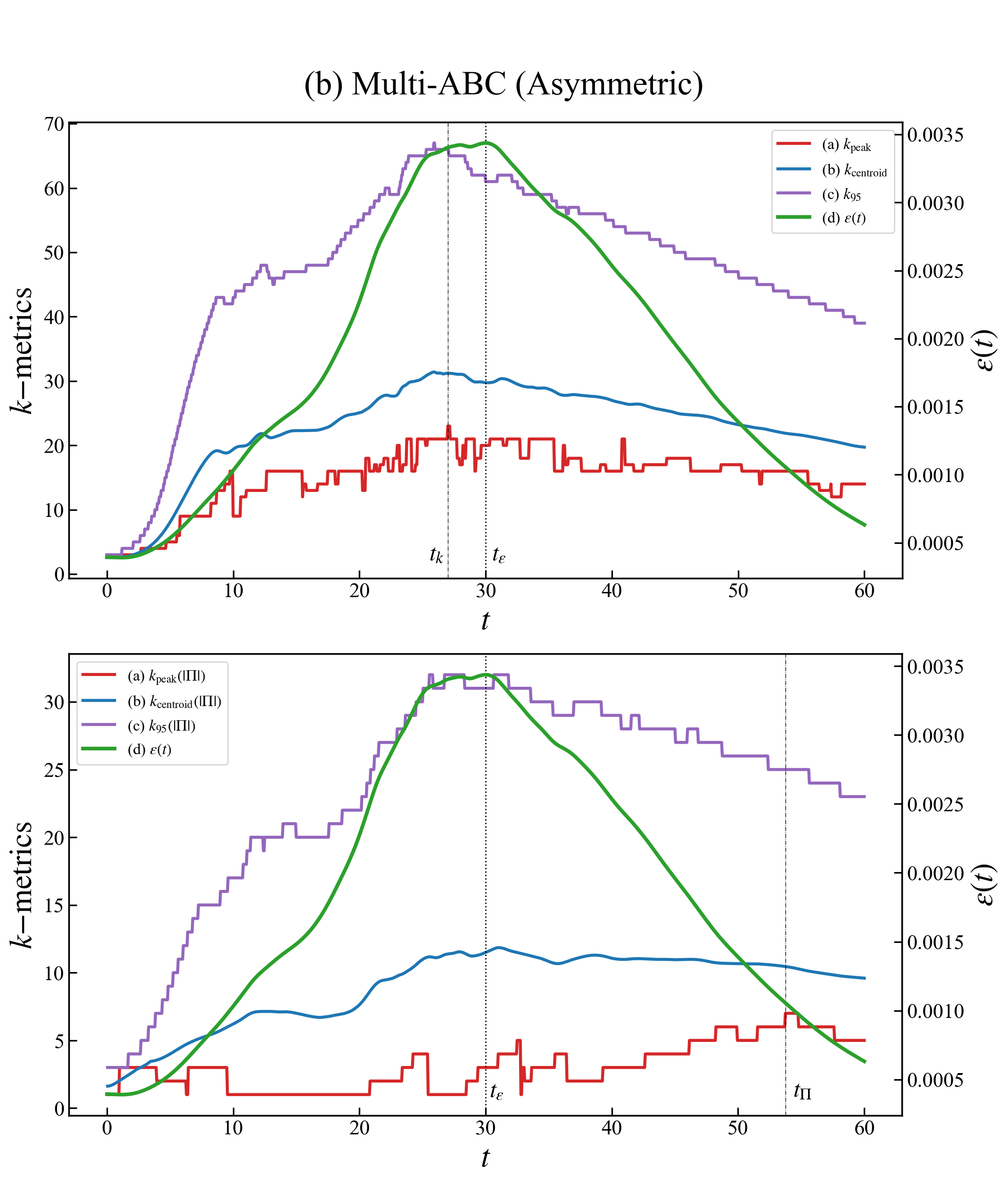}
  \caption{\textbf{Summary of peak/centroid/quantile wavenumber measures and their timing in (b) Multi-ABC (Asymmetric).}
The \emph{upper} panel shows characteristic wavenumber measures extracted from the curvature-weighted spectrum
\HL{$\Cfour(k,t)$} (equivalently $k^4\E(k)$): $k_{\mathrm{peak}}$ (red), $k_{\mathrm{centroid}}$ (blue), and $k_{95}$ (purple),
together with the dissipation rate $\varepsilon(t)$ (green, right axis).
The \emph{lower} panel shows the corresponding measures computed from the flux diagnostic \HL{$|\PiC(q)|$}
(denoted $k_{\mathrm{peak}}(\Pi)$, $k_{\mathrm{centroid}}(\Pi)$, and $k_{95}(\Pi)$), again with $\varepsilon(t)$ for reference.
Vertical markers indicate the characteristic times defined in Sec.~\ref{secIV:diagnostics}: \HL{$t_k=\arg\max_t\,k_{\mathrm{peak}}[\Cfour](t)$}, $t_\varepsilon=\arg\max_t\,\varepsilon(t)$, and $t_\Pi=\arg\max_t\,k_{\mathrm{peak}}[|\PiC|](t)$.
This combined view enables a direct comparison between (i) the early progression of curvature-weighted small-scale signatures and
(ii) the later evolution of interscale transfer, demonstrating the consistent temporal ordering $t_k<t_\varepsilon<t_\Pi$ across all initial conditions.}
\label{fig:peaks-variations-ABCMulti_Asymmetric}
\end{figure*}

\begin{figure*}[t]
  \centering
  \includegraphics[width=\linewidth]{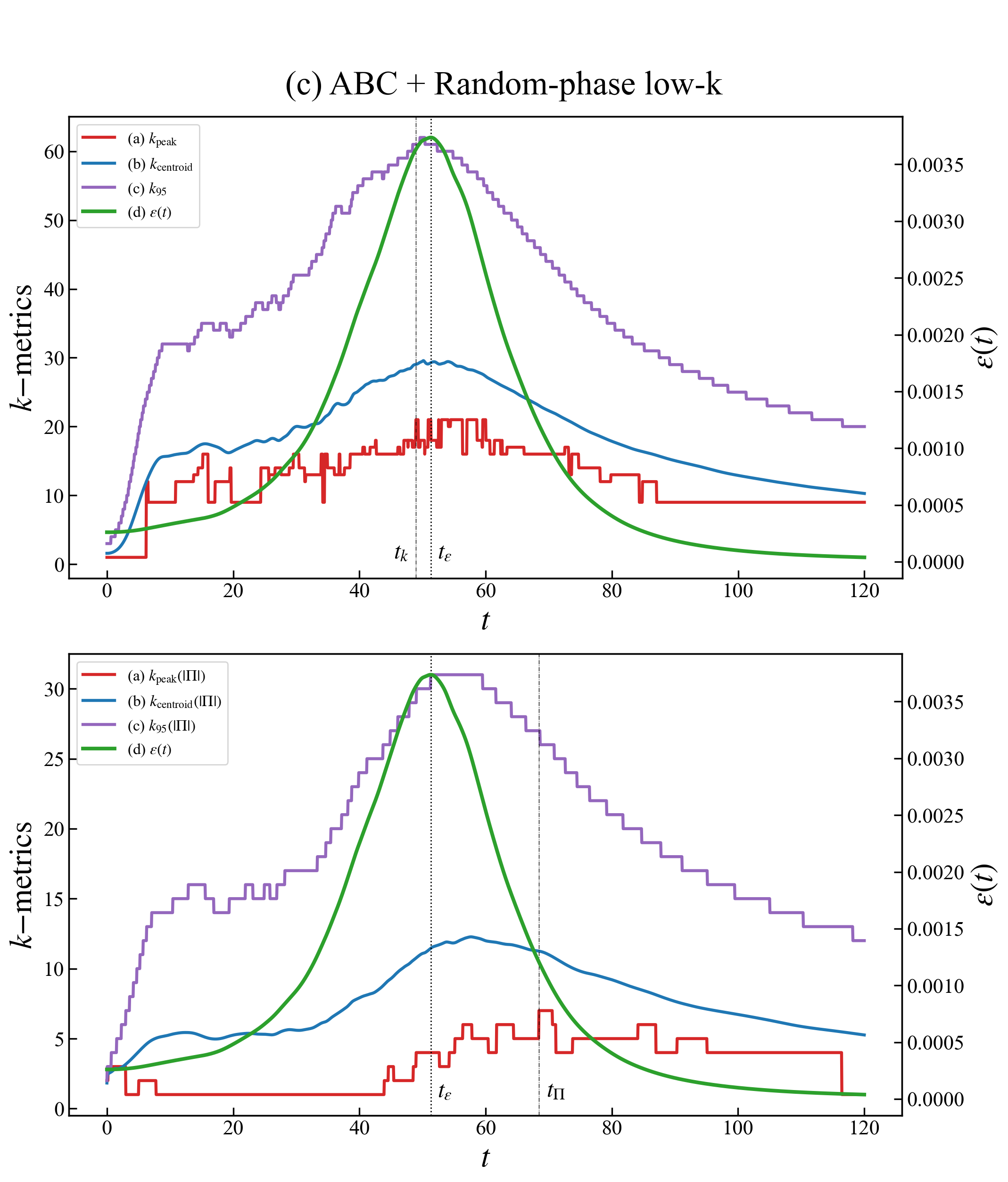}
  \caption{\textbf{Summary of peak/centroid/quantile wavenumber measures and their timing in (c) ABC + random-phase low-$k$.}
The \emph{upper} panel shows characteristic wavenumber measures extracted from the curvature-weighted spectrum
\HL{$\Cfour(k,t)$} (equivalently $k^4\E(k)$): $k_{\mathrm{peak}}$ (red), $k_{\mathrm{centroid}}$ (blue), and $k_{95}$ (purple),
together with the dissipation rate $\varepsilon(t)$ (green, right axis).
The \emph{lower} panel shows the corresponding measures computed from the flux diagnostic \HL{$|\PiC(q)|$}
(denoted $k_{\mathrm{peak}}(\Pi)$, $k_{\mathrm{centroid}}(\Pi)$, and $k_{95}(\Pi)$), again with $\varepsilon(t)$ for reference.
Vertical markers indicate the characteristic times defined in Sec.~\ref{secIV:diagnostics}: \HL{$t_k=\arg\max_t\,k_{\mathrm{peak}}[\Cfour](t)$}, $t_\varepsilon=\arg\max_t\,\varepsilon(t)$, and $t_\Pi=\arg\max_t\,k_{\mathrm{peak}}[|\PiC|](t)$.
This combined view enables a direct comparison between (i) the early progression of curvature-weighted small-scale signatures and
(ii) the later evolution of interscale transfer, demonstrating the consistent temporal ordering $t_k<t_\varepsilon<t_\Pi$ across all initial conditions.}
\label{fig:peaks-variations-ABC_RandomLowK}
\end{figure*}

\begin{figure*}[t]
  \centering
  \includegraphics[width=\linewidth]{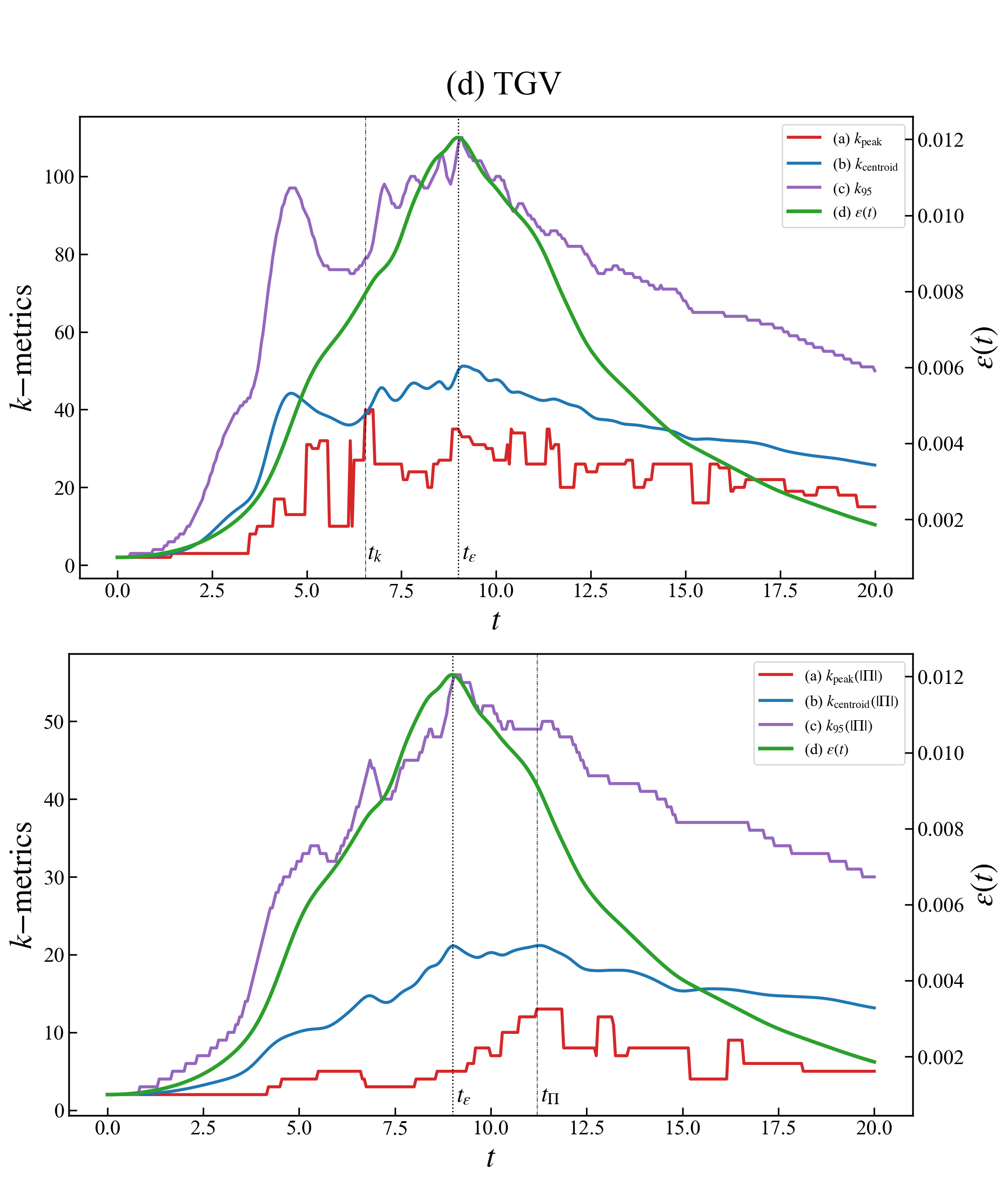}
  \caption{\textbf{Summary of peak/centroid/quantile wavenumber measures and their timing in (d) TGV.}
The \emph{upper} panel shows characteristic wavenumber measures extracted from the curvature-weighted spectrum
\HL{$\Cfour(k,t)$} (equivalently $k^4\E(k)$): $k_{\mathrm{peak}}$ (red), $k_{\mathrm{centroid}}$ (blue), and $k_{95}$ (purple),
together with the dissipation rate $\varepsilon(t)$ (green, right axis).
The \emph{lower} panel shows the corresponding measures computed from the flux diagnostic \HL{$|\PiC(q)|$}
(denoted $k_{\mathrm{peak}}(\Pi)$, $k_{\mathrm{centroid}}(\Pi)$, and $k_{95}(\Pi)$), again with $\varepsilon(t)$ for reference.
Vertical markers indicate the characteristic times defined in Sec.~\ref{secIV:diagnostics}: \HL{$t_k=\arg\max_t\,k_{\mathrm{peak}}[\Cfour](t)$}, $t_\varepsilon=\arg\max_t\,\varepsilon(t)$, and $t_\Pi=\arg\max_t\,k_{\mathrm{peak}}[|\PiC|](t)$.
This combined view enables a direct comparison between (i) the early progression of curvature-weighted small-scale signatures and
(ii) the later evolution of interscale transfer, demonstrating the consistent temporal ordering $t_k<t_\varepsilon<t_\Pi$ across all initial conditions.}
\label{fig:peaks-variations-TGV}
\end{figure*}

\begin{figure*}[t]
  \centering
  \includegraphics[width=\linewidth]{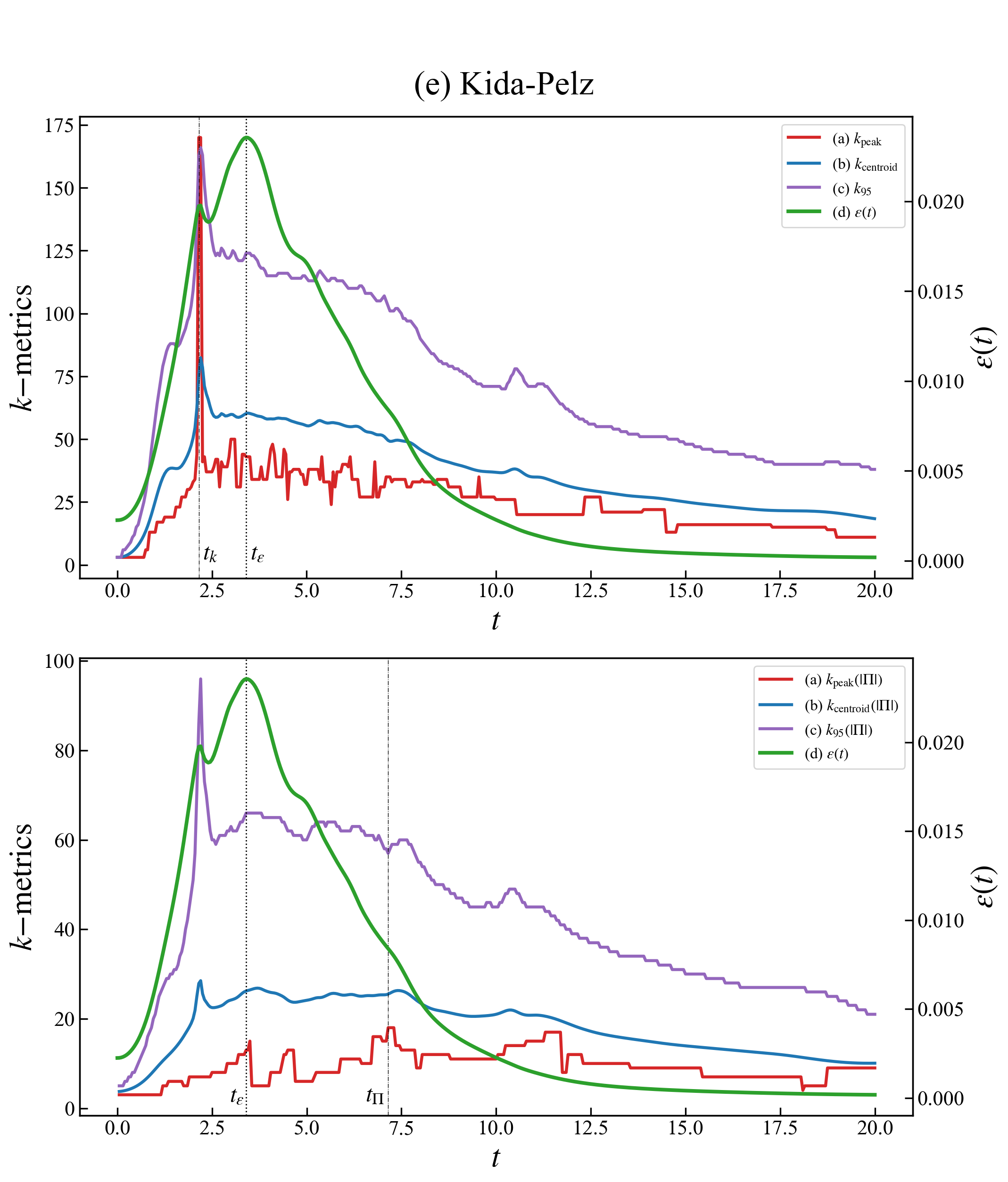}
  \caption{\textbf{Summary of peak/centroid/quantile wavenumber measures and their timing in (e) Kida--Pelz.}
The \emph{upper} panel shows characteristic wavenumber measures extracted from the curvature-weighted spectrum
\HL{$\Cfour(k,t)$} (equivalently $k^4\E(k)$): $k_{\mathrm{peak}}$ (red), $k_{\mathrm{centroid}}$ (blue), and $k_{95}$ (purple),
together with the dissipation rate $\varepsilon(t)$ (green, right axis).
The \emph{lower} panel shows the corresponding measures computed from the flux diagnostic \HL{$|\PiC(q)|$}
(denoted $k_{\mathrm{peak}}(\Pi)$, $k_{\mathrm{centroid}}(\Pi)$, and $k_{95}(\Pi)$), again with $\varepsilon(t)$ for reference.
Vertical markers indicate the characteristic times defined in Sec.~\ref{secIV:diagnostics}: \HL{$t_k=\arg\max_t\,k_{\mathrm{peak}}[\Cfour](t)$}, $t_\varepsilon=\arg\max_t\,\varepsilon(t)$, and $t_\Pi=\arg\max_t\,k_{\mathrm{peak}}[|\PiC|](t)$. In this $512^3$ case, the detected \HL{$k_{\mathrm{peak}}[\Cfour](t)$} can reach the analysis cutoff $k_{\max}^{(\mathrm{mask})}$ during the early transient (endpoint hit), reflecting cutoff-proximate peak locking; panel (f) ($1024^3$) shows the resolved reference behavior without such locking (cf.\ Fig.~\ref{fig:inspections}).}
\label{fig:peaks-variations-KidaPelz}
\end{figure*}

\begin{figure*}[t]
  \centering
  \includegraphics[width=\linewidth]{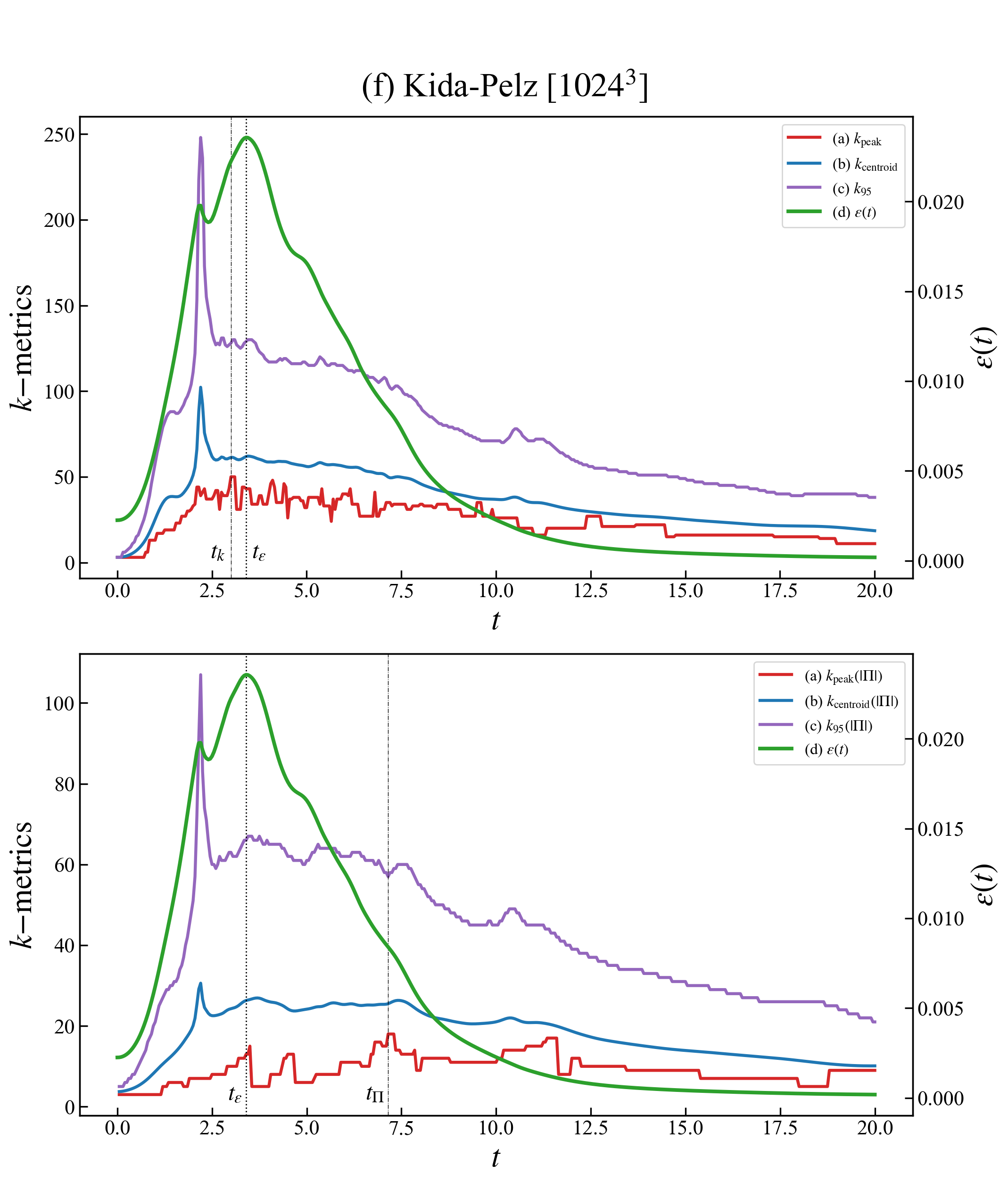}
  \caption{\textbf{Summary of peak/centroid/quantile wavenumber measures and their timing in (f) Kida--Pelz ${\rm [1024^3]}$.}
The \emph{upper} panel shows characteristic wavenumber measures extracted from the curvature-weighted spectrum
\HL{$\Cfour(k,t)$} (equivalently $k^4\E(k)$): $k_{\mathrm{peak}}$ (red), $k_{\mathrm{centroid}}$ (blue), and $k_{95}$ (purple),
together with the dissipation rate $\varepsilon(t)$ (green, right axis).
The \emph{lower} panel shows the corresponding measures computed from the flux diagnostic \HL{$|\PiC(q)|$}
(denoted $k_{\mathrm{peak}}(\Pi)$, $k_{\mathrm{centroid}}(\Pi)$, and $k_{95}(\Pi)$), again with $\varepsilon(t)$ for reference.
Vertical markers indicate the characteristic times defined in Sec.~\ref{secIV:diagnostics}: \HL{$t_k=\arg\max_t\,k_{\mathrm{peak}}[\Cfour](t)$}, $t_\varepsilon=\arg\max_t\,\varepsilon(t)$, and $t_\Pi=\arg\max_t\,k_{\mathrm{peak}}[|\PiC|](t)$.
This combined view enables a direct comparison between (i) the early progression of curvature-weighted small-scale signatures and
(ii) the later evolution of interscale transfer, demonstrating the consistent temporal ordering $t_k<t_\varepsilon<t_\Pi$ across all initial conditions.}
\label{fig:peaks-variations-KidaPelzHighRes}
\end{figure*}

We now present the main comparative results.
Figures~\ref{fig:peaks-variations-ABCMulti}--\ref{fig:peaks-variations-KidaPelzHighRes} show the time evolution of the characteristic wavenumber measures derived from the curvature-weighted spectrum \HL{$\Cfour(k,t)$} together with the dissipation rate $\varepsilon(t)$
(and, in the corresponding flux panels, the characteristic measures derived from \HL{$|\Pi(q,t)|$}).
The vertical markers indicate the characteristic times defined in Sec.~\ref{secIV:diagnostics}:
$t_k$ for the curvature-weighted precursor, $t_\varepsilon$ for the dissipation peak, and $t_\Pi$ for the
largest-flux-peak scale time.

Across all cases, the precursor time associated with the curvature-weighted spectrum occurs first:
\begin{equation}
t_k < t_\varepsilon,
\end{equation}
and the flux-related time occurs last,
\begin{equation}
t_\varepsilon < t_\Pi,
\end{equation}
thereby establishing the robust ordering
\begin{equation}
t_k < t_\varepsilon < t_\Pi
\end{equation}
beyond the TGV setting of Ref.~\cite{TsuzukiTGVPrecursorPRF}.
Importantly, this conclusion does not rely on a single, potentially noisy definition of a ``peak''.
Figures~\ref{fig:peaks-variations-ABCMulti}--\ref{fig:peaks-variations-KidaPelzHighRes} show that alternative wavenumber characterizations---including centroid-type and quantile-type measures (e.g., $k_c$ and $k_{95}$, and their associated characteristic times)---exhibit
the same qualitative advance relative to $t_\varepsilon$, providing evidence that the precursor is a robust
feature of the overall spectral redistribution rather than a fragile local maximizer.

The comparative view also clarifies how initial-condition-dependent transients modulate the magnitude of
the lead time while preserving the ordering.
In the ABC+random-phase low-$k$ case, the entire evolution is stretched, and both $t_k$ and $t_\varepsilon$
occur later than in the structured ABC-based cases; nevertheless, the curvature-weighted measures still rise
and stabilize before the dissipation maximum.
In the Kida--Pelz configuration, the precursor develops extremely early, consistent with the rapid formation
of tightly curved tubes and sheets seen in the $Q$-criterion visualizations (Sec.~\ref{secIII:InitCondSimSet}), and the high-resolution
run confirms that this early stabilization is a physical feature rather than an artifact of insufficient
small-scale resolution.

Finally, the flux-based characteristic measures evolve more slowly than their curvature-weighted counterparts,
reflecting the fact that $|\Pi|$ is an integral indicator of net interscale transfer and therefore lags the
initial formation of high-curvature structures that viscosity will soon act upon.
This systematic lag underlies the observed ordering $t_\varepsilon < t_\Pi$ across all cases.

\subsection{Summary of characteristic times}
\label{subsec:results_summarytable}

Table~\ref{tab:times_summary} summarizes, for each initial condition, the characteristic times
$t_k$, $t_\varepsilon$, and $t_\Pi$ extracted from the peak-time definitions introduced in
Sec.~\ref{secIV:diagnostics}, together with the corresponding lead and lag times,
\begin{equation}
\Delta t_{k\varepsilon} \equiv t_\varepsilon - t_k,
\qquad
\Delta t_{\varepsilon\Pi} \equiv t_\Pi - t_\varepsilon .
\end{equation}
The table makes the cross-case robustness of the temporal ordering
\begin{equation}
t_k < t_\varepsilon < t_\Pi
\end{equation}
immediately apparent, while also quantifying how the lead and lag magnitudes vary with
the initial condition.

In particular, $\Delta t_{k\varepsilon}$ provides a compact measure of how far in advance
the curvature-weighted diagnostic stabilizes relative to the dissipation peak,
whereas $\Delta t_{\varepsilon\Pi}$ quantifies the systematic delay of the flux-related
characteristic time.
Across all cases listed in Table~\ref{tab:times_summary}, the precursor time $t_k$
consistently precedes $t_\varepsilon$, and the flux-related time $t_\Pi$ occurs later,
demonstrating that the ordering $t_k < t_\varepsilon < t_\Pi$ is not restricted to a
single initial condition.

For the Kida--Pelz flow, only the $1024^3$ results are reported in Table~\ref{tab:times_summary}.
As discussed in Sec.~\ref{subsec:spike-inspection}, the $512^3$ simulation can exhibit
cutoff-proximate peak locking in the $k^4$-weighted spectrum during the early transient,
whereas the higher-resolution run provides a well-resolved peak away from the cutoff and a reliable
basis for extracting characteristic times.
The $512^3$ Kida--Pelz case is therefore excluded from the summary table and is used
only to illustrate resolution sensitivity and diagnostic considerations.

Additional robustness information based on alternative wavenumber measures,
including centroid- and quantile-type definitions (e.g., $k_{\mathrm{centroid}}$
and $k_{95}$ for both $|\curl\bfw|^2$ and $|\PiC|$), is provided in
all cases of Figs.~\ref{fig:peaks-variations-ABCMulti}--\ref{fig:peaks-variations-KidaPelzHighRes} and supports the same qualitative temporal ordering.

For visualization in the Q-snapshot sequences (Figs.~\ref{fig:SimSnap_ABCMulti_512}--\ref{fig:SimSnap_KidaPelz_512}), we annotated auxiliary reaching times
$t_c$ and $t_{95}$ based on global measures of the curvature-weighted spectrum
($k_{\rm centroid}$ and $k_{95}$), using the sustained-threshold definition in Eq.~(\ref{eq:reaching-time})
(with $q=0.9$ and $M=3$). Because Eq.~(\ref{eq:reaching-time}) is defined relative to
$k_m^\epsilon \equiv k_m(t_\epsilon)$, these reaching times are \emph{retrospective} by construction
and were not used in the primary ordering arguments.
Nevertheless, the relative timing of $(t_c,t_{95})$ with respect to $t_\epsilon$
provides a compact diagnostic of whether bulk redistribution toward small scales
approaches its $t_\epsilon$-level appreciably before, or only near, the dissipation peak.
Table~\ref{tab:aux_reaching_times} summarizes the annotated values from Figs.~\ref{fig:SimSnap_ABCMulti_512}--\ref{fig:SimSnap_KidaPelz_512}.
For the Kida--Pelz case, although Fig.~\ref{fig:SimSnap_KidaPelz_512} shows the $512^3$ visualization, we verified that the measured
values of $t_\epsilon$, $t_c$, and $t_{95}$ are identical in the companion $1024^3$ reference simulation
used elsewhere for the Kida--Pelz analysis (Sec.~\ref{subsec:spike-inspection} and Table~\ref{tab:times_summary}), so the values are representative of
the resolved Kida--Pelz dynamics.

\subsection{Viscosity dependence in Taylor--Green turbulence}
\label{subsec:results_viscosity}

The ordering $t_k<t_\varepsilon<t_\Pi$ established above pertains to the baseline viscosity $\nu=10^{-3}$.
To examine how this temporal ordering depends on the viscous damping of small scales, we repeat the TGV
simulations at two additional viscosities, $\nu=2.5\times 10^{-4}$ and $\nu=10^{-2}$, while keeping the
numerical scheme and the post-processing pipeline unchanged.

\begin{figure*}
\includegraphics[width=\linewidth]{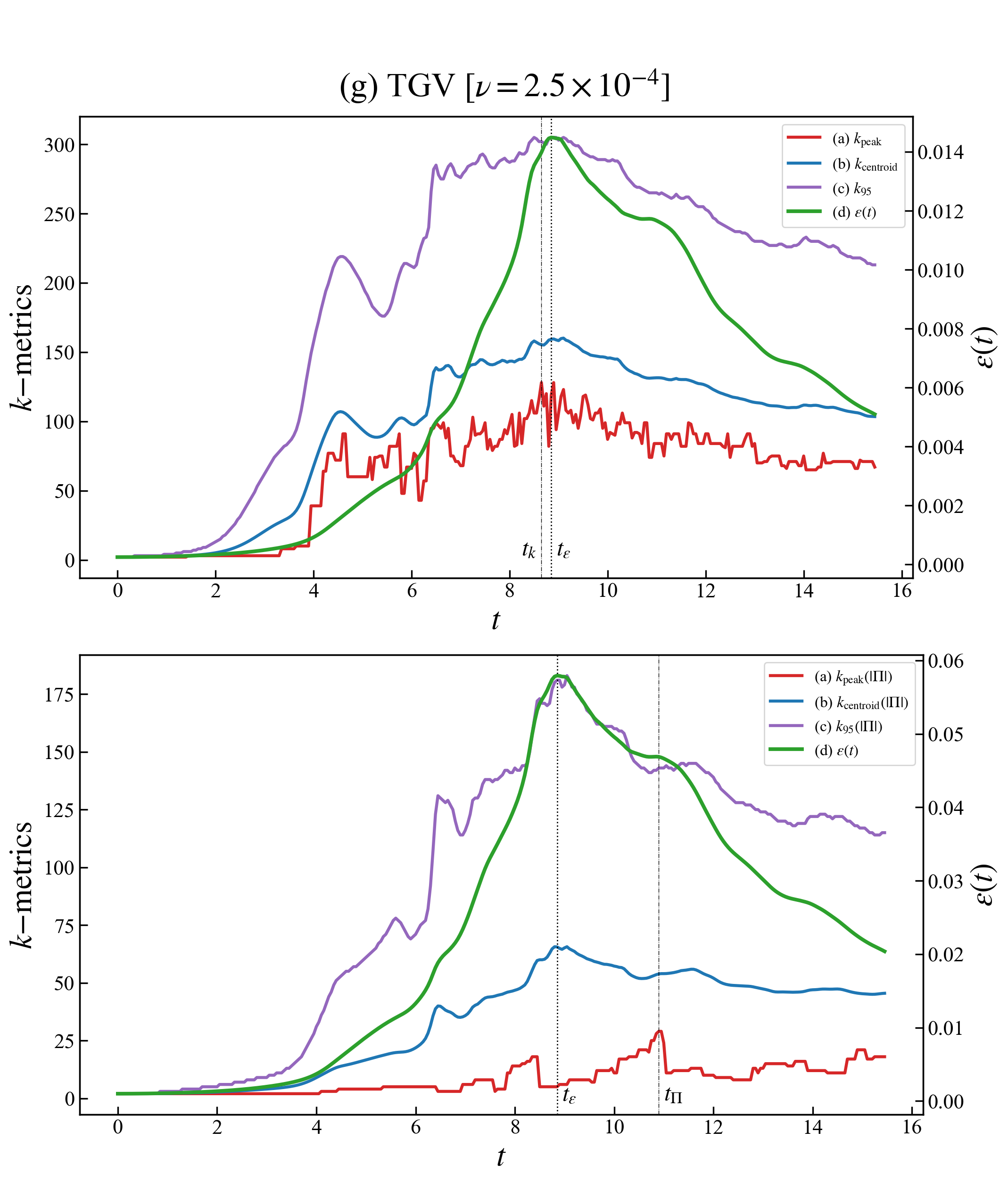}
\caption{
Viscosity variation in Taylor--Green vortex flow: low-viscosity case $\nu=2.5\times 10^{-4}$ (reference resolution $N^3=1024^3$).
The \emph{upper} panel shows the time series of peak-based wavenumber measures of the curvature-weighted spectrum \HL{$\Cfour(k,t)$}:
$k_{\mathrm{peak}}$, $k_{\mathrm{centroid}}$, and $k_{95}$, together with $\varepsilon(t)$.
Vertical markers indicate $t_k=\arg\max_t k_{\mathrm{peak}}(t)$ and the dissipation peak time $t_\varepsilon$.
The \emph{lower} panel shows the peak-based measures for the nonlinear energy-flux spectrum \HL{$|\PiC(q)|$}, with $t_\Pi=\arg\max_t k_{\mathrm{peak}}[|\PiC|](t)$. The ordering $t_k<t_\varepsilon<t_\Pi$ remains intact at this lower viscosity.}
\label{fig:visc-low-kpeaks}
\end{figure*}

\begin{figure*}
\includegraphics[width=\linewidth]{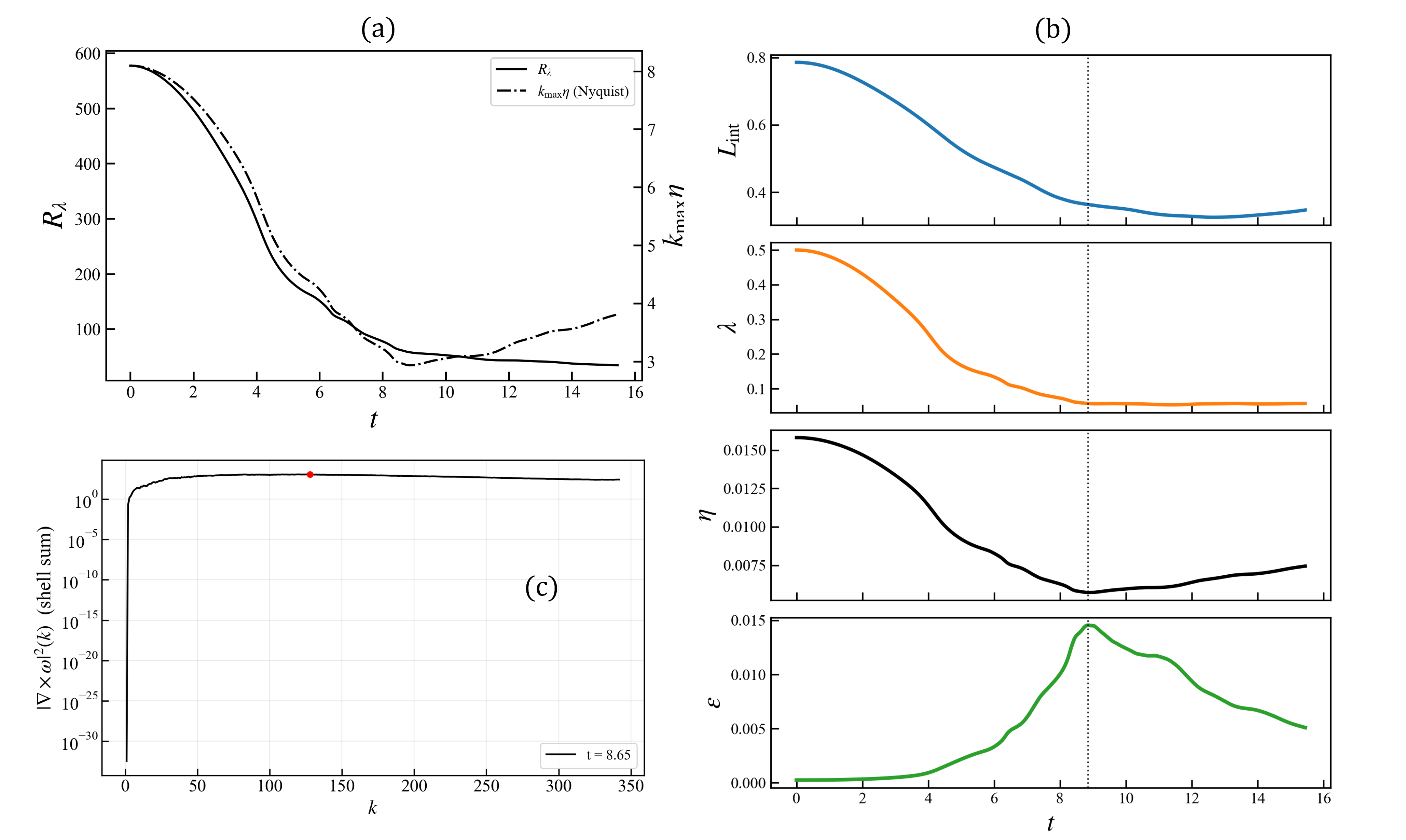}
\caption{
Viscosity variation in Taylor--Green vortex flow: low-viscosity case $\nu=2.5\times 10^{-4}$ (reference resolution $N^3=1024^3$).
(a) Resolution and Reynolds-number diagnostics $R_\lambda(t)$ and $k_{\max}\eta(t)$.
(b) Integral, Taylor, and Kolmogorov scales and the dissipation rate.
(c) Inspection of \HL{$\Cfour(k,t)$} at $t=t_k$, showing a well-defined peak away from the cutoff (cf.\ Sec.~\ref{subsec:results_viscosity}).}
\label{fig:visc-low-metrics}
\end{figure*}

Figure~\ref{fig:visc-low-kpeaks} shows the low-viscosity case $\nu=2.5\times 10^{-4}$.
At $512^3$, we observed (not shown) that the intensified small-scale activity, together with the $k^4$ weighting in \HL{$\Cfour(k,t)$}, drives the instantaneous argmax toward the analysis cutoff, leading to cutoff-proximate peak ``locking'' (cf.\ the baseline Kida--Pelz behavior discussed in Sec.~\ref{subsec:spike-inspection}).
We therefore treat the $512^3$ low-viscosity run as under-resolved for peak detection and use a $1024^3$
simulation as the reference for this case.
The spectrum inspection in Fig.~\ref{fig:visc-low-metrics}(c) confirms that the detected peak of \HL{$\Cfour(k,t)$} is
well separated from the cutoff at representative times.
With this adequately resolved dataset, the argmax-based peak scales again satisfy $t_k<t_\varepsilon<t_\Pi$.

\begin{figure*}
\includegraphics[width=\linewidth]{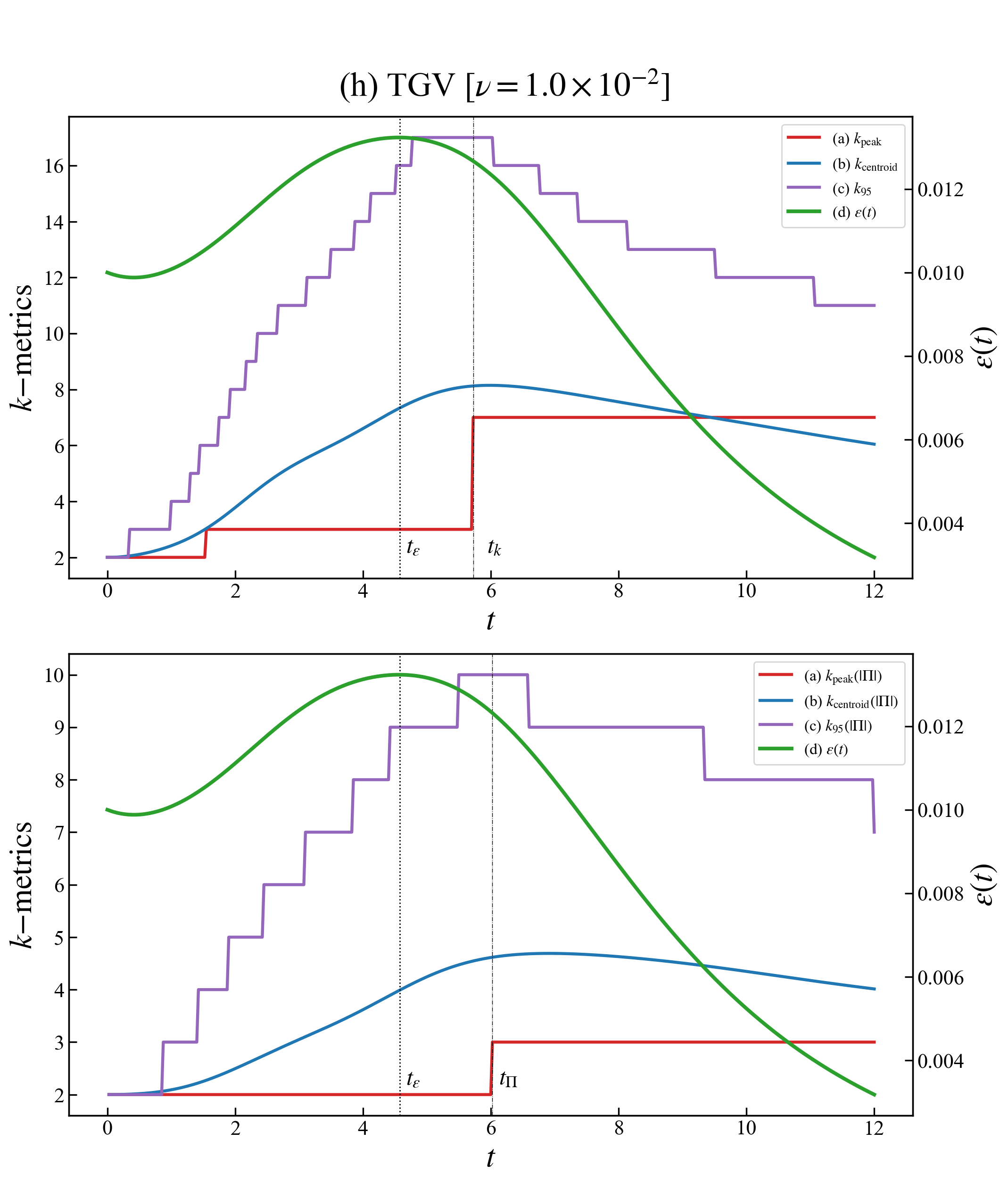}
\caption{
Viscosity variation in Taylor--Green vortex flow: high-viscosity case $\nu=10^{-2}$ ($N^3=512^3$).
Same layout as Fig.~\ref{fig:visc-low-kpeaks}.
\HL{In} this \HL{low-$R_\lambda$, limited-scale-separation case,} the dissipation peak precedes the maximum-attainment time of the curvature-weighted peak scale, $t_\varepsilon < t_k$, illustrating \HL{that} the \HL{precursor} ordering observed in the baseline cases \HL{need not persist when the accessible scale separation is limited}.}
\label{fig:visc-high-kpeaks}
\end{figure*}

\begin{figure*}
\includegraphics[width=\linewidth]{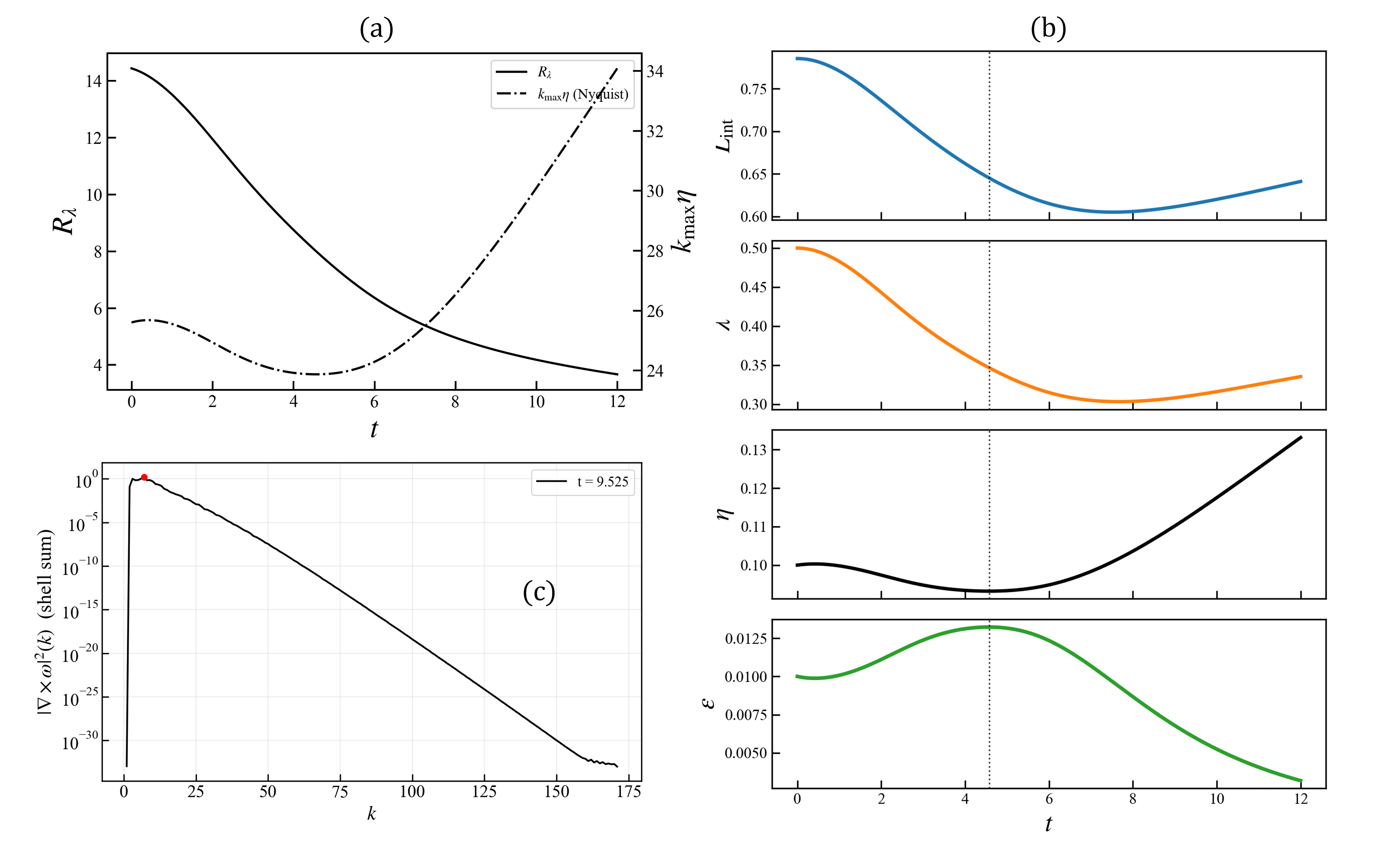}
\caption{
Viscosity variation in Taylor--Green vortex flow: high-viscosity case $\nu=10^{-2}$ ($N^3=512^3$).
Same layout as Fig.~\ref{fig:visc-low-metrics}.}
\label{fig:visc-high-metrics}
\end{figure*}

In contrast, the high-viscosity case $\nu=10^{-2}$ (Fig.~\ref{fig:visc-high-kpeaks}) shows that the temporal ordering
can change in a more strongly viscous, lower-Reynolds-number decay.
Here the spectrum remains comfortably separated from the cutoff (Fig.~\ref{fig:visc-high-metrics}(c)) and the resolution
indicator $k_{\max}\eta$ stays large (Fig.~\ref{fig:visc-high-metrics}(a)), so the change in ordering is not attributable
to high-$k$ locking.
Instead, the dissipation peak occurs earlier than the time at which the curvature-weighted peak scale reaches its
maximum, i.e.\ $t_\varepsilon < t_k$ (and $t_\Pi$ becomes comparable to $t_k$).

This breakdown is physically natural because the curvature-weighted spectrum is dominated by the highest resolved
wavenumbers and is therefore strongly affected by viscous damping.
In Fourier space the energy spectrum obeys
\begin{equation}
\partial_t \E(k,t)=T(k,t)-2\nu k^2 \E(k,t),
\label{eq:E_balance_for_C}
\end{equation}
so the curvature-weighted quantity $C(k,t)\equiv k^4\E(k,t)$ satisfies
\begin{equation}
\partial_t C(k,t) = k^4 T(k,t) -2\nu k^2 C(k,t),
\label{eq:C_balance}
\end{equation}
which shows that increasing $\nu$ accelerates the decay of high-$k$ contributions at a rate $\propto \nu k^2$.
\HL{Equations~\eqref{eq:E_balance_for_C}--\eqref{eq:C_balance} are used here as interpretive spectral-balance relations for the freely decaying problem. They do not imply ongoing external production at early times; all simulations are unforced after initialization.}
Consequently, when viscosity is large enough that the dissipative term dominates the high-$k$ balance, the early
advance of the peak scale observed at lower viscosity is weakened or can even reverse, reducing (or changing the
sign of) the lead time $\Delta t_{k\varepsilon}$.

These two additional TGV runs therefore indicate that the ordering $t_k<t_\varepsilon<t_\Pi$ is robust \HL{over the moderate-to-high-$R_\lambda$ cases examined here,} but \HL{need not persist in the low-$R_\lambda$/limited-scale-separation regime, where} the small-scale spectrum \HL{is strongly constrained and the precursor ordering can change. We therefore interpret the breakdown of the ordering in terms of Reynolds number and scale separation, rather than viscosity alone}.

\section{Discussion}\label{secVI:discussion}

\subsection{Physical interpretation of the ordering $t_k<t_\varepsilon<t_\Pi$}
\label{subsec:disc_interpretation}

The central finding of this study is that, for a broad class of freely decaying flows \HL{with sufficient Reynolds number and scale separation} (including all baseline cases at $\nu=10^{-3}$ and the low-viscosity TGV run at $\nu=2.5\times 10^{-4}$), the time $t_k$ at which the peak scale of the curvature-weighted spectrum \HL{$\Cfour(k,t)$} attains its maximum precedes the dissipation peak time $t_\varepsilon$, and $t_\varepsilon$ precedes the flux-related characteristic time $t_\Pi$, i.e., $t_k<t_\varepsilon<t_\Pi$. The additional high-viscosity TGV run at $\nu=10^{-2}$ (Sec.~\ref{subsec:results_viscosity}) provides a counterexample in which $t_\varepsilon<t_k$, highlighting that the ordering is a \HL{finite-Reynolds-number/scale-separation} feature rather than a kinematic identity.

Here we discuss a physical interpretation of this ordering in the context of non-stationary cascade development.

First, \HL{$\Cfour(k,t)$} is a curvature-biased measure: in incompressible flow $\curl\bfw=-\nabla^2\bfu$,
and therefore \HL{$\Cfour(k,t)\simeq k^4\E(k,t)$} strongly emphasizes the emergence of high-wavenumber content.
The advance of its characteristic wavenumber measures thus reflects the \emph{reach} of the flow toward
small scales in wavenumber space.
In a transient, freely decaying setting, the appearance of small scales in this sense can occur early,
driven by vortex stretching and folding and by the rapid steepening of gradients in localized regions.

However, the dissipation rate $\varepsilon(t)=2\nu\sum_k k^2\E(k,t)$ does not depend only on the \emph{reach} in $k$
but also on how much energy has accumulated in the dissipative range.
Even after the $k^4$-weighted spectrum has developed a stable peak scale,
the dissipative-range energy content can continue to grow for some time,
leading naturally to a later maximum of $\varepsilon(t)$.
This provides a mechanistic rationale for why $t_k<t_\varepsilon$ is expected in non-stationary transients.

Finally, the flux-based characteristic time $t_\Pi$ is extracted from \HL{$|\PiC(q)|$}, which is an integrated measure of net transfer across a scale.
Compared with \HL{$\Cfour(k,t)$}, the flux diagnostic is less localized in $k$ and reflects the
organization of triadic interactions across a range of scales \cite{Frisch1995,AlexakisBiferale2018}.
In a decaying transient, it is therefore plausible that the small-scale reach and the subsequent dissipation surge
precede the time at which the flux-related peak scale $k_{\mathrm{peak}}[|\PiC|](t)$ attains its maximum.
In this view, $t_\varepsilon<t_\Pi$ reflects that the dissipation maximum can occur while the cascade is still evolving toward its most pronounced interscale-transfer configuration.

\subsection{Initial-condition robustness and what varies across cases}
\label{subsec:disc_robustness}

A key motivation for the present work was to determine whether the precursor reported for TGV
is tied to its special symmetries or instead reflects a more general feature of small-scale formation in 3D turbulence.
Figures~\ref{fig:peaks-variations-ABCMulti}--\ref{fig:peaks-variations-KidaPelzHighRes} evidently show that the temporal ordering persists for
(i) structured helical ABC-based flows, (ii) a symmetry-broken multi-mode ABC superposition,
(iii) an ABC flow with additional random-phase low-$k$ content, and (iv) the high-symmetry Kida--Pelz configuration \cite{Kida1985,Boratav1994}.
This strongly suggests that the precursor is not specific to a particular topology or symmetry class.

While the ordering is robust, the \emph{timescales} vary substantially.
This is consistent with the fact that the route from initial organization to volume-filling turbulence depends on the
initial distribution of energy across large scales and on how quickly strong gradients develop.
For example, the ABC+random-phase low-$k$ case exhibits a stretched evolution in which both $t_k$ and $t_\varepsilon$
occur later than in the purely structured ABC-based cases, whereas the Kida--Pelz flow produces strong gradients very rapidly.
Such differences are also apparent in the evolution of global scales (Fig.~\ref{fig:scales-variations})
and in the $Q$-criterion visualizations (Sec.~\ref{secIII:InitCondSimSet}), which qualitatively indicate when tube-/sheet-dominated morphology becomes prevalent.

In this sense, $t_k$ acts as an \emph{early warning} marker whose absolute value is initial-condition dependent,
but whose relative position with respect to $t_\varepsilon$ and $t_\Pi$ appears to be a generic feature
within the parameter range studied here (fixed $\nu$, periodic box, freely decaying evolution).
\HL{To clarify whether this behavior is specific to the $p=4$ curl-of-vorticity/palinstrophy-related spectrum or is shared more broadly by high-wavenumber-weighted spectra, we compared the time evolution of $k_{\mathrm{peak}}$ for $p=4,6$, and $8$ for all initial conditions (Appendix Fig.~\ref{fig:appendix-highp}). Higher-order weights can also exhibit pre-$t_\varepsilon$ peak advancement, but they do not yield a systematically earlier or cleaner characteristic behavior. Instead, the peak trajectories generally become more irregular and more sensitive to the high-wavenumber tail as $p$ increases. We therefore interpret $p=4$ not as mathematically unique, but as a physically motivated and practically robust choice.}

\subsection{Why $k^4\E(k)$ is a sensitive precursor and why spike inspection is essential}
\label{subsec:disc_inspection}

The attractiveness of the curl-of-vorticity spectrum as a precursor lies in the strong $k^4$ weighting,
which amplifies the incipient development of small-scale content and makes the advance of a characteristic
wavenumber conspicuous already in the early transient. 
This sensitivity is what allows $t_{k}$ to precede $t_\varepsilon$ in the baseline and adequately scale-separated cases examined here.

At the same time, the same strong weighting makes peak-based metrics potentially vulnerable to
finite-resolution effects near the spectral cutoff.
A near-cutoff ``spike'' can arise from a combination of limited scale separation, discretization noise,
and the rapid growth of high-$k$ tails during intense transients.
Figure~\ref{fig:inspections} illustrates this issue most clearly for the Kida--Pelz initial condition at $512^3$,
where the naive argmax of \HL{$\Cfour(k,t)$} can approach the largest resolved wavenumbers.

This motivates explicit spike inspection (Sec.~\ref{subsec:spike-inspection}), together with a resolution check using $k_{\max}\eta(t)$ (Fig.~\ref{fig:kmaxeta-diagnostics}) and a higher-resolution reference for Kida--Pelz.
In this paper we do not ``correct'' peak locations by excluding a near-cutoff band; instead, cutoff-proximate peaks are used solely as a resolution diagnostic and motivate higher-resolution reference simulations.
In practice, the combination of (i) monitoring $k_{\max}\eta$, (ii) inspecting \HL{$\Cfour(k,t)$} at representative times,
and (iii) comparing peak-based measures with more global $k$-metrics such as $k_{\mathrm{centroid}}$ and $k_{95}$,
provides a robust workflow for deploying curvature-weighted diagnostics without conflating physical peak motion with cutoff artifacts.

An important practical implication is that spike inspection, while not standard for lower-order spectral measures,
is a natural and necessary component of analyses based on $k^4$-weighted spectra.
Including this check (Fig.~\ref{fig:inspections}) strengthens the credibility of the precursor claim,
particularly when comparing different initial conditions whose transient intensities differ markedly.

\subsection{Physical meaning of the curvature-weighted diagnostic}
\label{subsec:disc_physical_meaning}

In incompressible flow, \HL{$\nabla\cdot\bfu=0$ implies $\curl\bfw=-\nabla^2\bfu$}, so the spectrum \HL{$\Cfour(k,t)$} emphasized in this work is a spectral density of the Laplacian of the velocity, or \HL{equivalently a curvature-weighted velocity spectrum. Its physical meaning is most direct through palinstrophy. For periodic incompressible flow,}
\HL{
\begin{equation}
P(t) \equiv \frac12 \langle |\nabla\bfw|^2\rangle
= \frac12 \langle |\curl\bfw|^2\rangle,
\label{eq:palinstrophy_def}
\end{equation}
}
\HL{up to the standard normalization convention, and $\Cfour(k,t)$ is the corresponding shell-summed spectral density.
Palinstrophy appears} as the viscous sink in the enstrophy budget,
\HL{
\begin{equation}
\frac{d}{dt}\left(\frac12\langle |\bfw|^2\rangle\right)
= \langle \bfw\cdot\mathbf{S}\cdot\bfw\rangle
-\nu\langle |\nabla\bfw|^2\rangle,
\label{eq:enstrophy_budget}
\end{equation}
}
where \HL{$\mathbf{S}=(\nabla\bfu+\nabla\bfu^\top)/2$} is the strain-rate tensor.
\HL{The advance of $k_{\mathrm{peak}}[\Cfour](t)$ is therefore interpreted as a spectral marker} that vorticity gradients have reached \HL{dynamically active small scales before the total dissipative activity reaches its maximum. This} interpretation does not imply that \HL{viscous diffusion creates the} small scales; rather, nonlinear \HL{dynamics} creates steep gradients, and the curvature-weighted \HL{spectrum identifies the range} where viscous action will soon become \HL{important}.

\HL{We restrict the present interpretation to} classical incompressible Navier--Stokes turbulence. \HL{Although related differential operators may appear in other} continuum models, \HL{such models are not simulated or tested here and are outside the scope of the present conclusions}.

\subsection{Relation to transient cascade development and dissipation dynamics}
\label{subsec:disc_relation}

The present observations are consistent with a general picture of transient cascade development in decaying turbulence.
The early advance of the curvature-weighted spectral peak indicates that the flow reaches small scales in wavenumber space rapidly,
which aligns with the emergence of tightly curved tubes and rolled-up sheets visible in the $Q$-criterion snapshots (Sec.~\ref{secIII:InitCondSimSet}).
The subsequent dissipation peak reflects the accumulation of sufficient energy at dissipative scales,
and the later flux-related characteristic time reflects the continued evolution of interscale transfer organization.

From this perspective, the ordering $t_k<t_\varepsilon<t_\Pi$ can be interpreted as a temporal separation between
(i) the formation of small-scale \emph{reach} (geometry/curvature), (ii) the maximum of viscous activity (dissipation),
and (iii) the maturation of the most prominent transfer configuration captured by $|\PiC|$.
This separation is expected to be most visible in non-stationary transients where the cascade is not in equilibrium.
The fact that the ordering persists across initial conditions suggests that it is linked to generic properties of the nonlinear term
and to how gradients build up before they are fully dissipated.

\subsection{Limitations and outlook}
\label{subsec:disc_limitations}

The present study focuses on freely decaying turbulence in a periodic box at fixed viscosity $\nu=10^{-3}$,
with a base resolution of $512^3$ and an additional $1024^3$ reference for Kida--Pelz.
While this setup allows a controlled comparison across initial conditions,
several extensions would be valuable.

\HL{First, it remains to quantify how the lead time $\Delta t_{k\varepsilon}=t_\varepsilon-t_k$ depends on Reynolds number and scale separation.
Second, the present test set does not include homogeneous isotropic turbulence initialized from a statistically stationary forced state. Such a case would provide an additional canonical benchmark for assessing the generality of the precursor in a setting closer to fully developed isotropic turbulence, but it would require a separate statistically stationary precursor stage and is left for future work.
Third, all spectral diagnostics used in this work are isotropic shell sums. Consequently, the robustness reported here means robustness of the shell-averaged precursor across the tested decaying initial conditions; it does not imply that every direction-resolved or anisotropic spectral component follows the same timing. Direction-dependent spectra or angular-sector spectra would be useful for determining how the precursor is distributed among anisotropic modes, especially for highly structured initial conditions such as TGV, ABC, and Kida--Pelz flows. Such an anisotropy-resolved analysis is left for future work.
Finally, extensions to flows with additional physics (e.g., rotation, stratification, magnetohydrodynamics) would clarify the domain of applicability of curvature-weighted precursors.}

From a practical viewpoint, the precursor is most interesting in online or near-real-time settings. The argmax-based definitions of $t_k$ and $t_\Pi$ used in this paper are inherently retrospective because they require the future evolution to identify a global maximum. For online monitoring one could instead track a running maximum of \HL{$k_{\mathrm{peak}}[\Cfour](t)$} (and/or its growth rate) and trigger an ``alert time'' when the running maximum saturates (plateau detection) or when $k_{\mathrm{peak}}$ exceeds a prescribed threshold calibrated from reference datasets. A systematic evaluation of such online variants---including sensitivity to output cadence and the cutoff-spike inspection discussed in Sec.~\ref{subsec:spike-inspection}---is left for future work.

\begin{table*}
\caption{Summary of characteristic times for all initial conditions. Definitions follow Sec.~\ref{secIV:diagnostics}. 
All characteristic times are extracted on the output-time grid with spacing $\Delta t_{\mathrm{out}}=5\times10^{-2}$, 
implying a \HL{conservative} sampling uncertainty of \HL{up to $\Delta t_{\rm out}$} for each {characteristic} time and \HL{up to 2$\Delta t_{\rm out}$} out for the lead/lag measures.}

\label{tab:times_summary}
\begin{ruledtabular}
\begin{tabular}{lcccccc}
Case & $N^3$ & $t_k$ & $t_\varepsilon$ & $t_\Pi$ & $\Delta t_{k\varepsilon}$ & $\Delta t_{\varepsilon\Pi}$ \\
\hline
Multi-ABC & $512^3$ & 22.60 & 26.60 & 28.95 & 4.00 & 2.35 \\
Multi-ABC (Asymmetric) & $512^3$ & 27.00 & 30.00 & 53.75 & 3.00 & 23.75 \\
ABC + random-phase low-$k$ & $512^3$ & 48.95 & 51.30 & 68.45 & 2.35 & 17.15 \\
TGV & $512^3$ & 6.55 & 9.00 & 11.20 & 2.45 & 2.20 \\
Kida--Pelz (high resolution) & $1024^3$ & 3.00 & 3.40 & 7.15 & 0.40 & 3.75 \\
\end{tabular}
\end{ruledtabular}
\end{table*}

\begin{table*}[t]
\caption{
Auxiliary reaching-time markers $t_c$ and $t_{95}$ (Eq.~(\ref{eq:reaching-time}), with $q=0.9$ and $M=3$)
annotated in the Q-snapshot figures, and their lead/lag relative to the dissipation-peak time $t_\epsilon$.
We report $\Delta t_{c\epsilon}\equiv t_\epsilon-t_c$ and $\Delta t_{95\epsilon}\equiv t_\epsilon-t_{95}$
(positive values indicate \emph{lead} relative to $t_\epsilon$).
Because all times are sampled on the output-time grid with spacing $\Delta t_{\rm out}=5\times10^{-2}$,
differences with $|\Delta t|\le 2\Delta t_{\rm out}=0.1$ should be regarded as near-concurrent at the present cadence
(cf.\ Sec.~\ref{subsec:characteristic-times}).
For Kida--Pelz, the Q-snapshot annotations are taken from Fig.~\ref{fig:SimSnap_KidaPelz_512} ($512^3$ visualization), but the same values
($t_\epsilon$, $t_c$, and $t_{95}$) were confirmed in the companion $1024^3$ reference run (Sec.~\ref{subsec:spike-inspection}),
and thus can be interpreted on equal footing with the other cases. The peak-based characteristic-time summary in Table~\ref{tab:times_summary}
still uses the $1024^3$ Kida--Pelz reference dataset, consistent with our resolution-inspection protocol.
}
\label{tab:aux_reaching_times}
\begin{ruledtabular}
\begin{tabular}{lcccccc}
Case
& $t_\epsilon$
& $t_c$
& $t_{95}$
& $\Delta t_{c\epsilon}$
& $\Delta t_{95\epsilon}$
& Note
\\
\hline
Multi-ABC
& 26.60 & 27.00 & 26.85
& $-0.40$ & $-0.25$
& lag (fails)
\\
Multi-ABC (Asymmetric)
& 30.00 & 25.90 & 25.90
& $ 4.10$ & $ 4.10$
& lead (succeeds)
\\
ABC + random-phase low-$k$
& 51.30 & 50.15 & 49.60
& $ 1.15$ & $ 1.70$
& lead (succeeds)
\\
TGV
& 9.00 & 9.15 & 9.05
& $-0.15$ & $-0.05$
& near-concurrent at this cadence
\\
Kida--Pelz
& 3.40 & 2.20 & 2.20
& $ 1.20$ & $ 1.20$
& lead (succeeds; confirmed at $1024^3$)
\\
\end{tabular}
\end{ruledtabular}
\end{table*}

\paragraph*{Toward operational (online) early warning: an edge--bulk hybrid trigger.}
Table~\ref{tab:aux_reaching_times} shows that reaching-time markers based on global $k$-metrics
can provide an appreciable lead relative to the dissipation peak for several initial conditions
(including the Kida--Pelz case, for which $t_\epsilon$, $t_c$, and $t_{95}$ were confirmed to be identical
in the $1024^3$ reference run), but may be near-concurrent with, or even lag, $t_\epsilon$ in others.
This case dependence is consistent with the fact that centroid/quantile measures reflect bulk redistribution
and can be sensitive to non-monotonic adjustments when a sustained-threshold (``hold'') requirement is imposed
(Eq.~(\ref{eq:reaching-time})).
For operational monitoring, it is therefore natural to treat global metrics primarily as \emph{confirmatory}
or \emph{robustness} indicators, while using a leading-edge-sensitive diagnostic to provide the earliest alert.

To make this idea concrete on the discrete output grid $t_n=n\Delta t_{\rm out}$,
define \HL{$k_p(t_n)\equiv k_{\rm peak}[\Cfour](t_n)$} and the running maximum
\begin{equation}
k_p^{\max}(t_n)\equiv \max_{0\le m\le n} k_p(t_m).
\end{equation}
An online analogue of the ``advance-and-saturation'' notion is the first time at which the running maximum
has not increased over the past $M$ outputs:
\begin{equation}
t_{\rm sat}^{(p)} \equiv \min\left\{t_n:\; k_p^{\max}(t_n)=k_p^{\max}(t_{n-M})\right\},
\label{eq:tsat_online}
\end{equation}
which can be evaluated sequentially with a detection latency of $M\Delta t_{\rm out}$.
To reduce false alarms (e.g.\ cutoff-proximate peak locking), one may combine this edge trigger with
a bulk confirmation based on the centroid growth rate,
\HL{
\begin{equation}
g_c(t_n)\equiv \frac{k_{\rm centroid}[\Cfour](t_n)-k_{\rm centroid}[\Cfour](t_{n-M})}
{M\Delta t_{\rm out}},
\end{equation}
}
and a simple cutoff-proximity guard,
\begin{equation}
\mathcal{L}(t_n)\equiv \mathbf{1}\!\left[k_p(t_n)\ge \chi\, k_{\max}^{(\mathrm{mask})}\right],
\end{equation}
where $k_{\max}^{(\mathrm{mask})}$ is defined in Eq.~(\ref{eq:maskmax}) and $\chi\in(0,1)$ (e.g.\ $\chi=0.9$) sets a conservative margin.
A minimal hybrid alert time can then be defined as
\begin{equation}
t_{\rm alert}\equiv
\min\left\{t_n\ge t_{\rm sat}^{(p)}:\; g_c(t_n)\ge g_{c,\mathrm{th}}\ \ \wedge\ \ \mathcal{L}(t_n)=0\right\},
\label{eq:hybrid_alert}
\end{equation}
with a tunable threshold $g_{c,\mathrm{th}}$.
Systematically evaluating such online variants---including sensitivity to $(M,\chi,g_{c,\mathrm{th}})$ and to the output cadence
$\Delta t_{\rm out}$, and benchmarking across initial conditions and viscosity---is an important direction for future work.

\section{Conclusions}\label{secVII:conclusions}

We investigated spectral precursors to the dissipation peak in freely decaying, three-dimensional turbulence
by extending our recent PRF Letter on the Taylor--Green vortex (TGV) \cite{TsuzukiTGVPrecursorPRF}
to a broader set of initial conditions.
Using pseudo-spectral direct numerical simulations in a $2\pi$-periodic box at $\nu=10^{-3}$,
we considered five distinct $512^3$ runs (Multi-ABC, asymmetric Multi-ABC, ABC with random-phase low-$k$ perturbations,
TGV, and Kida--Pelz) and an additional $1024^3$ reference run for Kida--Pelz.
All data were analyzed using a unified post-processing pipeline for isotropic spectra, transfer, and flux diagnostics.

Our main conclusion is that the characteristic time associated with the advance of the curvature-weighted diagnostic
\HL{$\Cfour(k,t)\simeq k^4\E(k,t)$} robustly precedes the dissipation peak time, and that the dissipation peak precedes the
flux-related characteristic time extracted from \HL{$|\PiC(q)|$}.
Across all tested initial conditions we consistently observe the temporal ordering
\begin{equation}
t_k \;<\; t_\varepsilon \;<\; t_\Pi,
\end{equation}
thereby demonstrating that the precursor behavior reported previously for TGV is not restricted to that highly symmetric benchmark.
Moreover, alternative wavenumber measures based on centroid and quantile definitions (e.g., $k_{\mathrm{centroid}}$ and $k_{95}$)
exhibit the same qualitative early-warning behavior, supporting the robustness of the precursor beyond a single peak-based metric.

\HL{The comparison with $p=6$ and $p=8$ (Appendix Fig.~\ref{fig:appendix-highp}) shows that precursor-like peak advancement is not mathematically unique to the exponent $p=4$. Rather, high-wavenumber weighting more generally sensitizes spectra to the early formation of small scales. The practical significance of $p=4$ is that it is directly tied to the curl-of-vorticity/palinstrophy spectrum and, among the tested weights, provides a stable compromise between early responsiveness and finite-resolution robustness. The present results therefore support $\Cfour(k,t)$ as a physically interpretable diagnostic for the deterministic decaying-flow cases examined here, while leaving broader statistical, anisotropic, and forced-flow generalizations for future work.}

A notable methodological outcome is that curvature-weighted spectra require explicit spike inspection for cutoff-proximate artifacts.
Because \HL{$\Cfour(k,t)$} strongly emphasizes high wavenumbers, naive peak extraction can be contaminated when the argmax is influenced
by near-cutoff behavior, as illustrated for the $512^3$ Kida--Pelz case.
We therefore combined (i) resolution diagnostics using $k_{\max}\eta(t)$, (ii) direct inspections of \HL{$\Cfour(k,t)$}, and (iii) higher-resolution reference simulations where required (Kida--Pelz and the low-viscosity TGV case).
These procedures ensure that the extracted characteristic times reflect genuine small-scale dynamics rather than finite-resolution locking.

The results presented here suggest that curvature-weighted spectral diagnostics provide a practical and \HL{potentially useful} early indicator of impending dissipation activity in \HL{the} non-stationary turbulent transients \HL{tested here}.
In addition to their conceptual value for understanding transient cascade development,
these diagnostics may enable operational uses in large-scale simulations, such as scheduling dense output around the dissipation episode
or triggering adaptive resolution strategies.

Several directions remain open.
The viscosity-variation TGV runs in Sec.~\ref{subsec:results_viscosity} provide a first indication of Reynolds-number \HL{and scale-separation} dependence: the \HL{precursor ordering remains clear when the decay develops a sufficiently extended small-scale range, but it need not persist once the accessible scale separation becomes too limited}.
A systematic map of the lead time $\Delta t_{k\varepsilon}=t_\varepsilon-t_k$ versus Reynolds number, viscosity, and initial spectral content remains an important direction, as does testing the precursor in statistically stationary forced turbulence and exploring extensions to anisotropic or multi-physics settings.
These studies will help delineate the generality and practical scope of curvature-weighted precursors in turbulent flows.

\begin{figure*}[t]
  \centering
  \includegraphics[width=\linewidth]{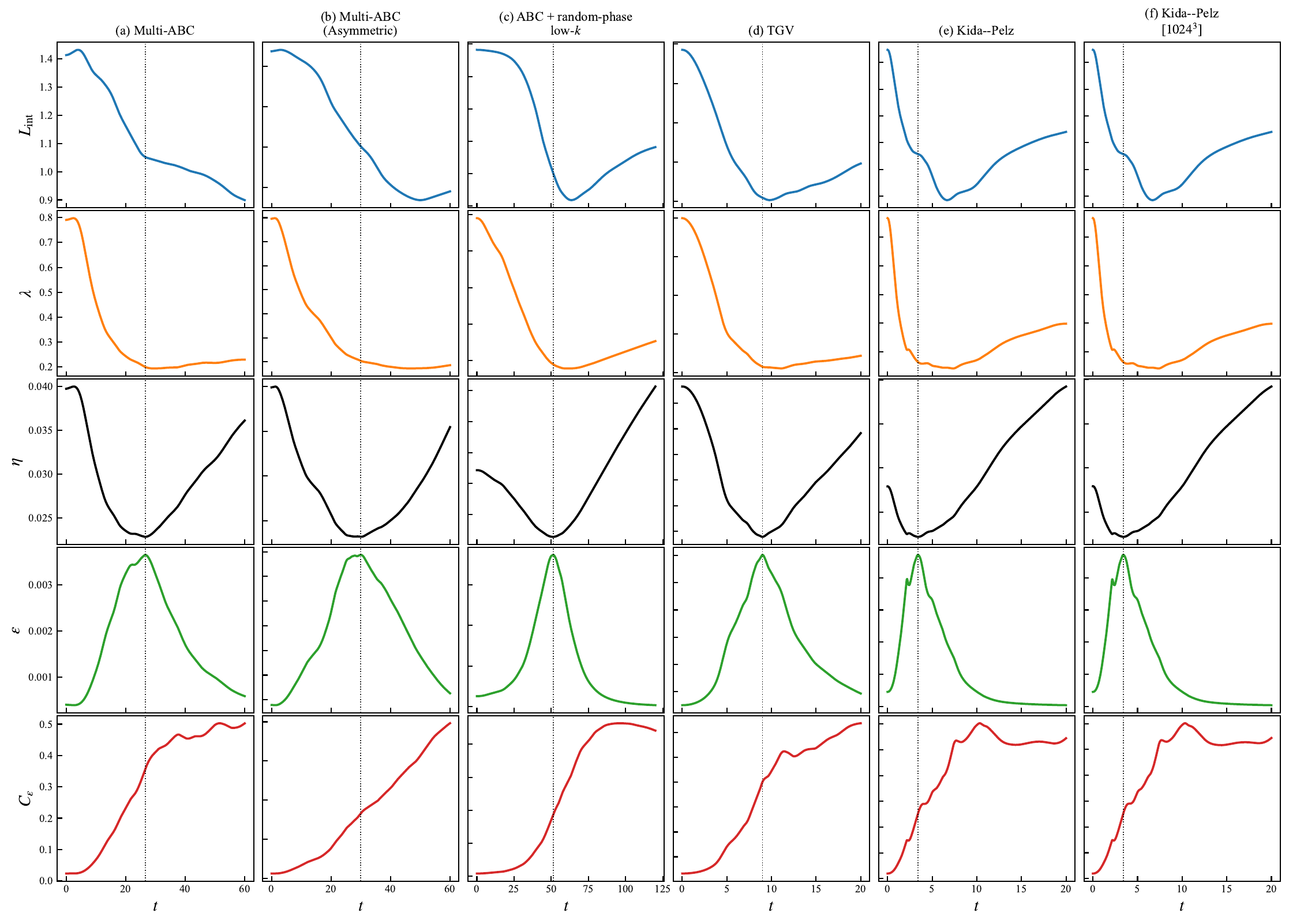}
  \caption{Time evolution of \HL{the} integral scale \HL{$L_{\mathrm{int}}$}, Taylor microscale \HL{$\lambda$}, Kolmogorov \HL{scale $\eta$}, dissipation rate \HL{$\varepsilon$}, \HL{and normalized dissipation coefficient $C_{\varepsilon}=\varepsilon L_{\mathrm{int}}/u'^3$} for \HL{the decaying flows studied here. Panels correspond to} (a) Multi-ABC, (b) Multi-ABC (Asymmetric), (c) ABC+random-phase low-$k$, (d) TGV, (e) Kida--Pelz ($512^3$), and (f) Kida--Pelz ($1024^3$). \HL{The vertical dashed line in each column marks} the dissipation-peak time \HL{$t_{\varepsilon}$}.}
\label{fig:scales-variations}
\end{figure*}

\appendix
\renewcommand{\thefigure}{A\arabic{figure}}
\setcounter{figure}{0}

\section{\HL{Additional high-wavenumber-weighted and velocity-spectrum diagnostics}}
\label{app:additional-spectra}

\subsection{\HL{Comparison across high-wavenumber weights}}
\label{app:highp-comparison}

\HL{To test whether the observed precursor behavior is specific to the curl-of-vorticity spectrum or is shared more broadly by high-wavenumber-weighted spectra, we compare $k_{\mathrm{peak}}$ for $p=4,6$, and $8$ in Appendix Fig.~\ref{fig:appendix-highp}. For $p=4$, the plotted spectrum is exactly the shell-summed curl-of-vorticity/palinstrophy-related spectrum $\Cfour(k,t)$ used in the main text. For $p=6$ and $p=8$, we use shell-center comparison spectra}
\HL{
\begin{equation}
\Cp(k,t)=k^{p-4}\Cfour(k,t), \qquad p=6,8,
\label{eq:appendix_Cp}
\end{equation}
}
\HL{so that the $p=4$ curve remains the same diagnostic as $k_{\mathrm{peak}}[\Cfour](t)$. The spectra $\Cp$ for $p=6,8$ are introduced only as diagnostic high-wavenumber-weighting comparisons and should be distinguished from the modal definition of $\Cfour(k,t)$ in Eq.~\eqref{eq:curlw2_equiv_methods}. The comparison shows that higher-order weights can also exhibit pre-$t_\varepsilon$ peak advancement, but increasing $p$ does not systematically improve the precursor behavior and generally leads to noisier, more high-$k$-sensitive peak trajectories.}

\subsection{\HL{Representative velocity spectra}}
\label{app:velocity-spectra}

\begin{figure*}[t]
  \centering
  \includegraphics[width=\linewidth]{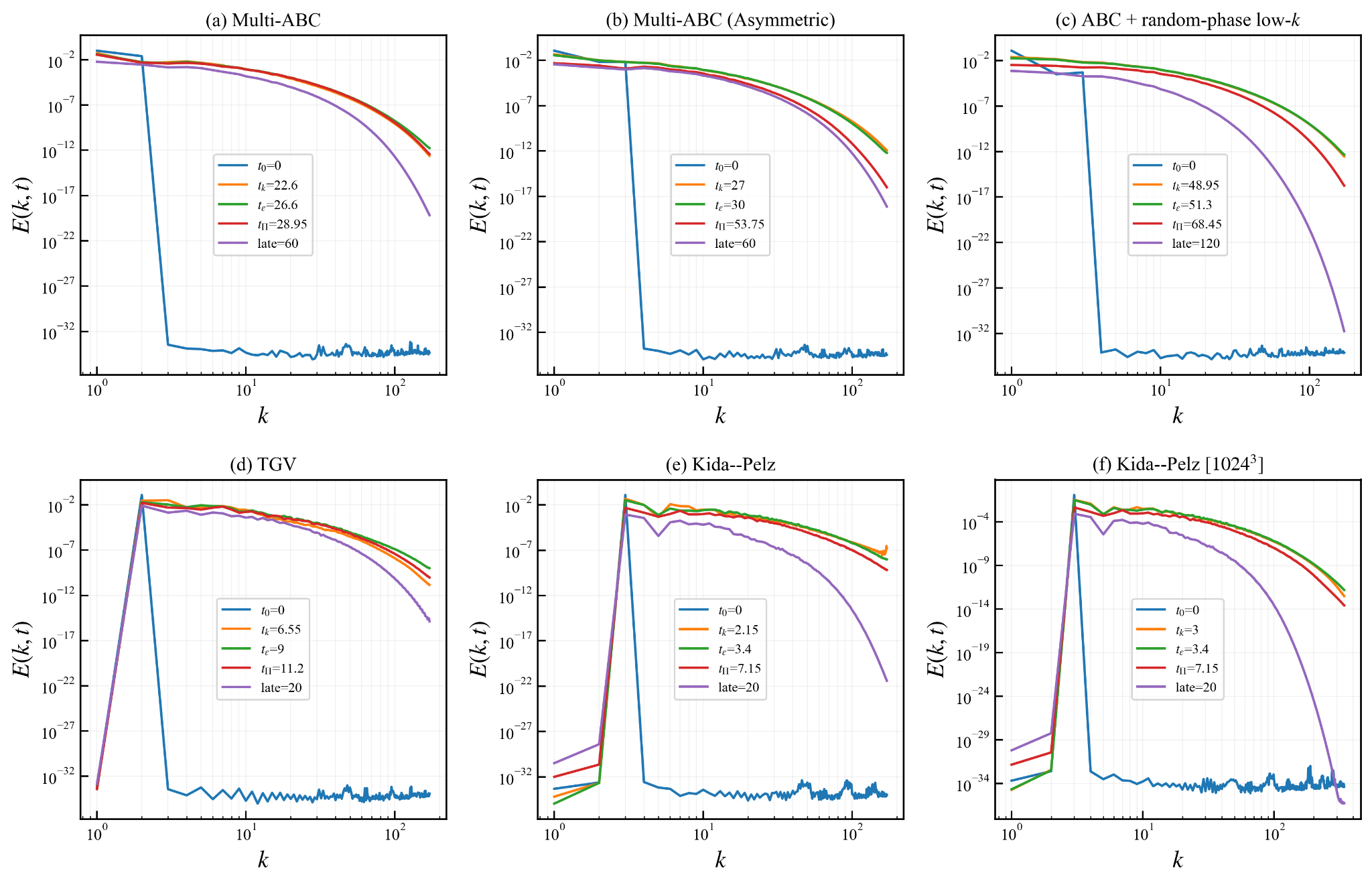}
  \caption{\HL{Representative velocity energy spectra $E(k,t)$ for the six simulations considered in Fig.~\ref{fig:scales-variations}: (a) Multi-ABC, (b) Multi-ABC (Asymmetric), (c) ABC+random-phase low-$k$, (d) TGV, (e) Kida--Pelz at $512^3$, and (f) Kida--Pelz at $1024^3$. In each panel, spectra are shown at the initial time, at the curvature-weighted precursor time $t_k$, at the dissipation-peak time $t_\varepsilon$, at the flux-related time $t_\Pi$, and at a late-time snapshot. The spectra are multiplied by the same calibration factor used in the time-series diagnostics, so that the initial total kinetic energy is normalized consistently across cases. Only shells retained by the spherical two-thirds analysis mask are shown. The vertical ranges are chosen independently in each panel to emphasize the spectral shape and temporal broadening within each case. This figure complements Fig.~\ref{fig:scales-variations} by showing the underlying redistribution of kinetic energy toward higher wavenumbers during the transient decay.}}
  \label{fig:appendix-velocity-spectra}
\end{figure*}

\HL{For each initial condition, Appendix Fig.~\ref{fig:appendix-velocity-spectra} shows representative velocity spectra $E(k,t)$ at selected times during the evolution. These spectra complement Fig.~\ref{fig:scales-variations} by showing the underlying redistribution of kinetic energy toward higher wavenumbers during the transient decay.}

\begin{figure*}[t]
  \centering
  \includegraphics[width=\linewidth]{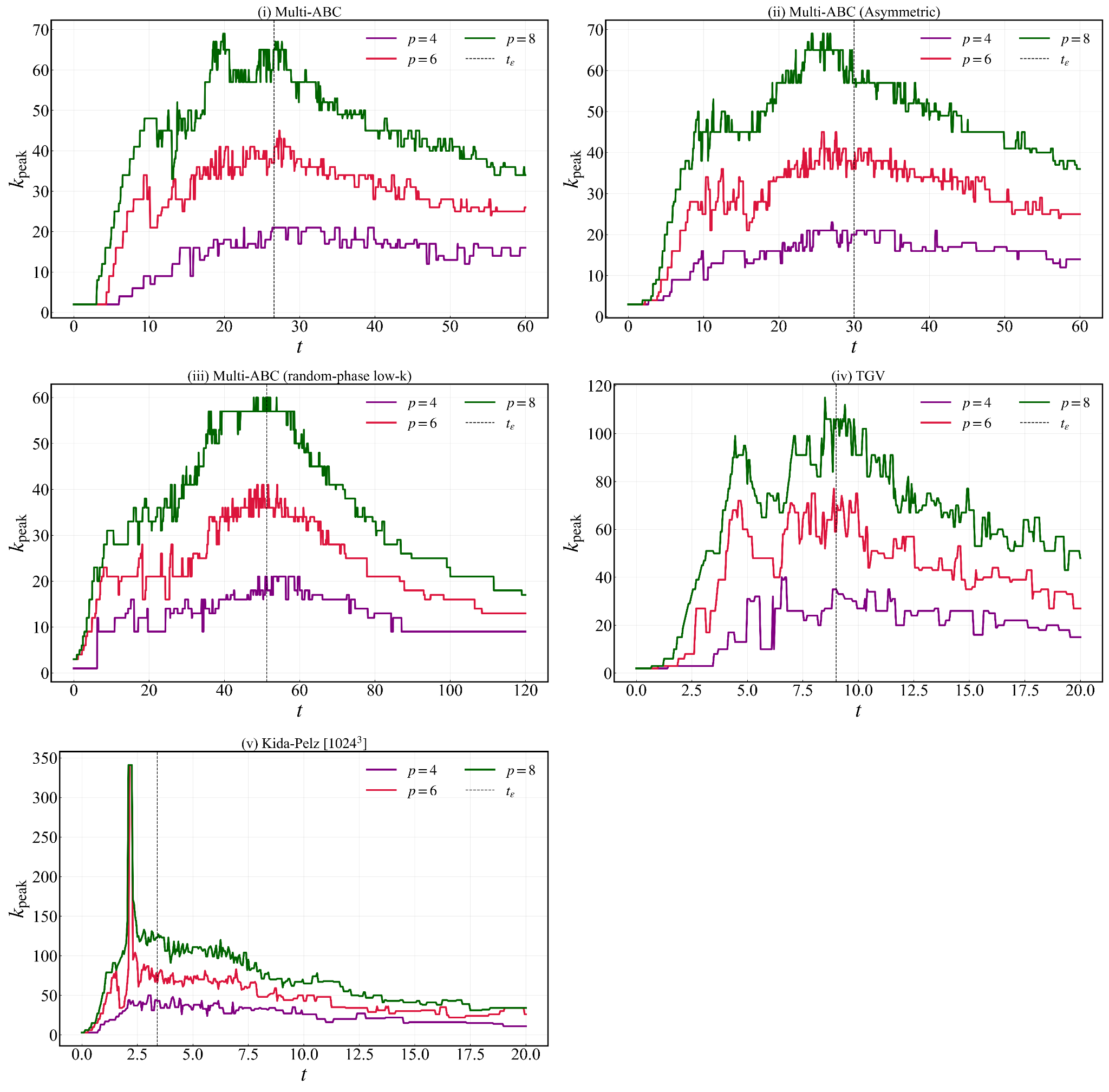}
  \caption{\HL{Time evolution of the peak wavenumber $k_{\mathrm{peak}}$ of the high-wavenumber-weighted spectra for $p=4,6,8$ for the five initial conditions considered in the present study. The dashed vertical line marks the dissipation-peak time $t_\varepsilon$. For $p=4$, the plotted quantity corresponds to the shell-summed curl-of-vorticity/palinstrophy-related spectrum used in the main text; for $p=6$ and $p=8$, the comparison spectra are defined by $\Cp(k,t)=k^{p-4}\Cfour(k,t)$. The figure shows that higher-order weights can also exhibit pre-$t_\varepsilon$ peak advancement, but increasing $p$ does not systematically improve the precursor behavior and generally leads to noisier, more high-$k$-sensitive peak trajectories. This supports the use of $p=4$ as a physically interpretable and practically robust compromise.}}
  \label{fig:appendix-highp}
\end{figure*}

\paragraph*{Acknowledgments.}
This study was supported by JSPS KAKENHI (Grant Number 22K14177) and JST PRESTO (Grant Number JPMJPR23O7).

\paragraph*{Funding.}
JSPS KAKENHI (Grant Number 22K14177) and JST PRESTO (Grant Number JPMJPR23O7).

\bibliography{Manuscript}

\end{document}